\documentclass[a4paper,11pt]{article}
\pdfoutput=1
\usepackage{jheppub}
\usepackage{amsfonts,amssymb,amsmath}
\usepackage[normalem]{ulem}
\usepackage{color}
\usepackage{epstopdf}
\usepackage{tensor}
\usepackage{enumerate}
\usepackage{graphicx}
\usepackage{epsfig}
\usepackage{amsfonts,amssymb,amsmath,bm}
\usepackage{enumitem}
\usepackage[hang]{subfigure}
\usepackage{epstopdf}
\usepackage{mathtools}
\usepackage{xcolor}
\usepackage{comment}
\usepackage{mathtools}
\usepackage{soul}
\usepackage{calc}
\usepackage{soul}
\usepackage{comment}
\usepackage[utf8]{inputenc}

\newsavebox\CBox
\newcommand\hcancel[2][0.5pt]{%
  \ifmmode\sbox\CBox{$#2$}\else\sbox\CBox{#2}\fi%
  \makebox[0pt][l]{\usebox\CBox}%
  \rule[0.5\ht\CBox-#1/2]{\wd\CBox}{#1}}

\usepackage{tikz}
\usetikzlibrary{decorations.pathmorphing}
\usetikzlibrary{calc}

\newcommand{\be}{\begin{equation}}
\newcommand{\ee}{\end{equation}}
\newcommand{\ba}{\begin{eqnarray}}
\newcommand{\ea}{\end{eqnarray}}
\newcommand{\beal}{\begin{aligned}}
\newcommand{\eeal}{\end{aligned}}
\newcommand{\nn}{\nonumber}  
\newcommand{\beq}{\begin{equation}}
\newcommand{\eeq}{\end{equation}}

\title{Extended Thermodynamics and Complexity in Gravitational Chern-Simons Theory}

\author[a,b]{Antonia M. Frassino}
\author[c,d]{Robert B.~Mann,}
\author[e]{Jonas R. Mureika}

\affiliation[a]{Departament de F{\'\i}sica Qu\`antica i Astrof\'{\i}sica, Institut de
Ci\`encies del Cosmos,\\ Universitat de
Barcelona, Mart\'{\i} i Franqu\`es 1, E-08028 Barcelona, Spain}
\affiliation[b]{ D{\'e}partement de Physique Th{\'e}orique and Center for Astroparticle Physics,\\
Universit{\'e} de Gen{\`e}ve, 24 quai Ansermet, CH–1211 Gen{\`e}ve 4, Switzerland}
\affiliation[c]{ Perimeter Institute for Theoretical Physics, 31 Caroline St. N.,Waterloo,\\ Ontario N2L 2Y5, Canada}
\affiliation[d]{ Department of Physics and Astronomy, University of Waterloo,\\ Waterloo, Ontario N2L 3G1, Canada}
\affiliation[e]{\it Department of Physics, Loyola Marymount University, Los Angeles,\\ California~USA~~90045}

\emailAdd{antoniam.frassino@icc.ub.edu}
\emailAdd{rbmann@uwaterloo.ca}
\emailAdd{jmureika@lmu.edu}

\abstract{We study  several aspects of the extended thermodynamics of   BTZ black holes with thermodynamic mass  $M=\alpha m + \gamma \frac{j}{\ell}$ and angular momentum  $J = \alpha j + \gamma \ell m$, for general values of the parameters 
$(\alpha, \gamma)$ ranging from regular ($\alpha=1, \gamma=0$) to exotic ($\alpha=0, \gamma=1$).
  We show that there exist two distinct behaviours for the black holes, one when $\alpha > \gamma$ (``mostly regular''), and the other when $\gamma < \alpha$ (``mostly exotic'').  We find that the Smarr formula  holds 
 for all $(\alpha, \gamma)$. We derive the corresponding thermodynamic volumes, which we find to be positive provided $\alpha$ and $\gamma$ satisfy a certain  constraint.  The dependence of pressure on volume is unremarkable and strictly decreasing when $\alpha > \gamma$, but a maximum volume emerges for large $J\gg T$ when $\gamma > \alpha$; consequently an exotic black hole of a given horizon circumference and temperature can exist in two distinct anti de Sitter backgrounds. 
  We compute the reverse isoperimetric ratio, and study the Gibbs free energy and criticality conditions for each.  {Finally we investigate the complexity growth of these objects and find that they are all proportional to the complexity of the BTZ black hole.  Somewhat surprisingly, purely exotic BTZ black holes have vanishing complexity growth.  }
  }

\keywords{Black Holes, Thermodynamics, Volume, AdS/CFT correspondence, Complexity, BTZ}
\date{} 

\begin{document}

\maketitle

\section{Introduction}

Black hole physics and thermodynamics have an interesting relationship reaching back over  more than
four decades  \cite{Bekenstein:1973ur,hawking,fourlaws}, one that remains an active research area of considerable import in theoretical physics, primarily due to its implications for quantum gravity.  It is particularly relevant  in  spacetimes with  a negative cosmological constant as it provides connections between otherwise distinct theoretical concepts in the context of the AdS/CFT correspondence conjecture \cite{Maldacena1998}, which relates string theories (and therefore gravitation) formulated on asymptotically-anti de Sitter (AdS)  spacetimes to a conformal field theory on the spacetime boundary.

The cosmological constant, $\Lambda$, has generally been regarded as an inert parameter in black hole thermodynamics, one that sets a new length scale but does little else in this context.  In recent years, however, the role of $\Lambda$ has been found to be quite significant, and a new sub-discipline called   ``black hole chemistry'' has emerged \cite{Kubiznak:2014zwa}.  In this approach, each thermodynamic parameter in an asymptotically AdS black hole has a chemical counterpart in representations of the first law, with 
$\Lambda$ associated with pressure.  From a cosmological perspective a negative cosmological constant induces a vacuum pressure, and so the former association is quite natural.  A new $PV$ term is thereby introduced into the first law, where $V$ is the volume conjugate to $P$.  The presence of a black hole in the spacetime can be regarded as a displacement of vacuum energy, leading to an interpretation of the black hole mass $M$ as a gravitational analogue of chemical enthalpy \cite{Kubiznak:2014zwa,KastorEtal:2009}.  

Subsequent investigation has indicated that  black hole thermodynamics has a very close correspondence with real world systems \cite{KastorEtal:2009, Dolan:2010, Dolan:2011a, Dolan:2011b} including Van der Waals fluids \cite{KubiznakMann:2012}, reentrant phase transitions \cite{Altamirano:2013ane,GunasekaranEtal:2012,Frassino:2014pha}, triple points analogous to the triple point in water \cite{Altamirano:2013uqa} and nuclear matter systems \cite{Frassino:2017htb}.  More recently a holographic interpretation of some of these phenomena has been 
studied \cite{Karch:2015rpa,Sinamuli:2015drn,Sinamuli:2017rhp}, and
it has been shown that  black holes can be regarded as holographic heat engines, where  renormalization group flow performs the cycles \cite{Johnson:2014yja,Caceres:2015vsa,Nguyen:2015wfa,Hennigar:2017apu}.  Very recently a class of hairy black holes was discovered that exhibit superfluid phase transitions analogous to those of superfluid helium \cite{Hennigar:2016xwd}.  A thorough review describing these phenomena and much more has recently appeared \cite{Kubiznak:2016qmn}.

One of the first insights of black hole chemistry is that in order to satisfy the Smarr relation \cite{Smarr:1972kt}, it is necessary to treat  $\Lambda$ as thermodynamic pressure.  In any dimension the
first law and its associated  Smarr relation are
\ba
d(G_D M)&=&TdS+\Omega dJ+VdP \label{TDD} \\
(D-3)G_D M&=&(D-2)TS+(D-2)\Omega J-2VP     + (D-3)\Phi Q  
\label{relationsD}
\ea
 for a charged singly-rotating black hole, 
where $J$ is its angular momentum, $\Omega$ its angular velocity, $T$ its temperature
and $S$ its entropy,  and the $D$-dimensional Newton constant $G_D$ has been explicitly retained.  The thermodynamic pressure $P$ is 
\beq
\label{pressD}
P=-\frac{\Lambda}{8\pi} =  \frac{(D-2)(D-1)}{16\pi l^2}
\eeq
and its conjugate volume is $V$.  Despite factors of $(D-3)$ and $(D-2)$ appearing in the above formulae, it has recently been shown that \eqref{relationsD} holds in both the $D \to 3$ and $D\to 2$ limits
\cite{Frassino:2015oca}.   Although there were no phase transitions in either dimension, the
 $D=3$ case led to a number of interesting findings.

In particular,  the definition of thermodynamic volume  was somewhat subtle, with the volume  depending on both  horizon size and charge, quite unlike
 the situation for charged black holes in higher dimensions  \cite{KubiznakMann:2012,GunasekaranEtal:2012}.  Conversely, retaining a definition of volume depending only on horizon size necessitates introduction of a new work term associated with the renormalization scale of the black hole mass, in turn modifying both the Smarr formula \eqref{relationsD}  and first law \eqref{TDD} in $D=3$.
 
Our interest in this paper is in better understanding of the $D=3$ case in the presence of a gravitational Chern-Simons term in the three dimensional action. We  consequently can discuss   features of the so-called ``exotic" black hole solution.
These are black holes whose metric is of the same form as the $D=3$ BTZ black hole but whose parameters of mass and charge are interchanged.  These solutions were first obtained by considering the behaviour of topological matter in $D=3$ gravity \cite{Carlip:1994hq}.  Such matter does not couple to the spacetime metric but rather only to the connection.  The net effect is that the (negative) cosmological constant becomes a constant
of integration associated with the topological matter, and the conserved charges of mass and angular momentum become interchanged with one another.  Even stranger was that the entropy of these black holes is proportional to the area of their inner horizons \cite{Carlip:1994hq}.  
This situation can occur in other theories of gravity in $D=3$ \cite{Banados:1998dc}, and it was recently shown that the entropy of such exotic black holes still has a statistical interpretation \cite{Townsend:2013ela}.

In this paper we consider the chemistry of these exotic black holes.  We are particularly interested in 
understanding the   generalized first law  \eqref{TDD} and Smarr formula \eqref{relationsD} for these
objects, along with their thermodynamic volume and its properties.  We find a number of interesting distinctions
between these two cases.  The Gibbs free energy is in general always larger for exotic black holes than for their
standard BTZ counterparts. Furthermore, we find that in certain regions of parameter space an exotic black hole
can have two distinct pressures for the same thermodynamic volume.  Since volume is proportional to radius, this means that  an exotic black hole of a given horizon circumference and temperature can exist in two distinct AdS backgrounds. These black holes have a maximum size at a given temperature. Finally, we find that the volume of  a sufficiently small and sufficiently exotic black hole  violates the Reverse Isoperimetric Inequality \cite{Cvetic:2010jb}; only for $\alpha \geq \gamma$ is this inequality always satisfied.  This means that some exotic black holes, though having lower entropy than their BTZ counterparts, contain more entropy than their thermodynamic volumes would otherwise naively allow.

Moreover, we study the late time complexity growth for the exotic BTZ black hole using the framework of the conjectured equality Complexity$=$Action, where the action is calculated in the Wheeler de Witt (WdW) patch. Unlike previous work \cite{Cai:2016xho}, we take into account the corner terms
\cite{Booth:2001gx,Lehner:2016vdi}.
We find that the growth rate for the standard BTZ black hole is as expected, but (somewhat surprisingly) that  exotic black holes have no additional contributions to the growth rate. This provides further
evidence that the  complexity = action conjecture depends on which action one takes into account
\cite{Brown:2018bms,Goto:2018iay}.

\section{ 
Black hole solution from gravitational Chern-Simons}

The Einstein-AdS action in $D=3$ is
\begin{equation}
I_{EH} = \frac{1}{8\pi G_N} \int_{M^3} \left[  e_{a}\wedge R^{a}-\frac{1}{6\ell^{2}}\epsilon^{abc}%
e_{a}\wedge e_{b}\wedge e_{c}\right]  
\label{3fNI}
\end{equation}
in the well-known Einstein-Cartan formulation, 
where 
\begin{equation}
T^{a}=de^{a}+\epsilon^{abc}\omega_{b}\wedge e_{c}\,,\qquad R^{a}=d\omega^{a}+\frac
{1}{2}\epsilon^{abc}\omega_{b}\wedge\omega_{c}
\label{torsion}
\end{equation}
are the respective torsion and curvature forms,  with dreibein
1-forms $e^{a}$ and Lorentz connection 1-forms $\omega^{a}$ ($a=0,1,2$), and the exterior product of forms is implicit. 

The field equations have the exact solution
\begin{equation}\label{BTZ1}
    ds^2 = - N^2 dt^2 + N^{-2} dr^2 + r^2 \left( d\phi + N^{\phi} dt \right)^2
\end{equation}
with the lapse and the angular shift function equal to 
\begin{equation}\label{lpseshft}
    N^2 = -8 G_N m + \frac{r^2}{\ell^2} + \frac{16 G_N^2 j^2}{r^2},\,\,\,
    N^{\phi} = -\frac{4 G_N j}{r^2}\,.    
\end{equation}
where $\ell$ is the $AdS_3$ radius, $G_N$ is the $\left( 2+1 \right)$-dimensional Newton constant and $m$ and $j$ are constants of integration, whose interpretation is contingent upon the theory under consideration.

The simplest gravity model for which BTZ black holes are exotic is the 
parity-odd  action for $D=3$ Einstein gravity with $\Lambda < 0$  \cite{Witten:1988hc}. 
The action is
\begin{equation}
I_{GCS} = \frac{1 }{8\pi G_N}\int_{M^3} \left[ {\tilde{\ell}}  \omega_{a}\wedge\left(  d\omega^{a}+
 {\frac{1}{3}}
\epsilon^{abc}\omega_{b}\wedge\omega_{c}\right)  -\frac{1}{{ \hat{\ell}}}e_{a}\wedge
T^{a}\right]    \label{3fE}%
\end{equation}
whose parity transform is minus itself (and so its field equations preserve parity). {The quantities
$\hat{\ell}$ and $\tilde{\ell}$ are coupling constants whose values need not be equal.}
 Varying with respect to $e^{a}$ yields $T^{a}=0$, which in turn allows one to solve for
$\omega^{a}$.  Inserting the result into the equation obtained by variation with respect to $\omega^a$
gives  the same field equations as those of the (parity-even) Einstein-AdS action (\ref{3fNI}),
{with $\ell$ replaced by the geometric mean $\sqrt{\hat{\ell}\tilde{\ell}}$.}

An alternative action for obtaining the exotic BTZ black hole is \cite{Carlip:1994hq}
\be
I_{BCE\omega} = 
\frac{\ell}{8\pi G_N}\int_{M^3} \left(e^a\wedge R_a[\omega]+B^a\wedge D_\omega C_a\right) 
\label{bceaaction}
\ee
where each of $\{e^a, \omega^a, B^a, C^a\}$ are regarded as independent fields, with 
$D_\omega$   the covariant derivative with respect to the connection $\omega^a$.  Note that there is no coupling of matter to the metric in \eqref{bceaaction}; for this reason the fields $\{B^a, C^a\}$ are referred to as topological matter. A variety of solutions exist to the field equations of this theory, one of which is the black hole metric 
\eqref{BTZ1} \cite{Carlip:1994hq}.

 Like a typical rotating black hole solution in $D=4$, the horizon equation for a BTZ black hole ($N=0$) has two solutions that correspond to two Killing horizons, located respectively at
\begin{equation}
\label{rpm}
    r_{\pm} = 2 \sqrt{G_N \ell \left(\ell m \pm \sqrt{\ell^2 m^2-j^2}\right)}
\end{equation}
that give the condition $\ell m >|j|$ (necessary to have the two event horizons) and define the extremal case when $\ell m = |j|$. Assuming that $j>0$ then the parameters $m$ and $j$ of the black hole can be expressed as \cite{Miskovic:2009kr}
\begin{equation}
    m = \frac{r_{-}^2+r_{+}^2}{8 G_N \ell^2},\;\;j = \frac{r_{-} r_{+}}{4 G_N \ell} \label{mANDj}
\end{equation}

However $m$ and $j$ in general are not respectively proportional to the mass and angular momentum of the black hole. This was first pointed out for the action 
\eqref{bceaaction}, where a computation of the Noether charges indicated that $M \propto j$ and $J \propto m$,
with $M$ and $J$ the respective conserved mass and angular momentum \cite{Carlip:1994hq}.  
{Setting $ \tilde{\ell} =\hat{\ell} = \ell$ for simplicity,}
more recently it has been noted that  \cite{Miskovic:2009kr,Townsend:2013ela}
 \begin{eqnarray}
     M &=& \alpha m  +\gamma \frac{j}{\ell}\,, \label{SystM}\\
     J &=&  \alpha j + \gamma \ell m  \,. \label{SystJ}
\end{eqnarray}
can more generally be considered, with  $\left(\alpha,\,\gamma \right)=\left(1,\,0 \right)$ corresponding to the normal BTZ black hole, while the case $\left(\alpha,\,\gamma \right)=\left(0,\,1 \right)$ corresponds to the exotic BTZ black hole.  
We shall refer to the class of black holes with general $\left(\alpha,\,\gamma \right)$ as generalized exotic black holes. An analysis of the mass formula for the exotic BTZ black hole in the context of  weak cosmic censorship  is discussed in \cite{Zhang:2016tpm}.

We  note here that a possible action that could extrapolate between the standard and the exotic solutions could be defined as the sum of   two actions: the standard Einstein-Hilbert  action (EH) and the Gravitational Chern-Simons action (GCS), as used in \cite{Solodukhin:2005ah},
where the parameter in front of the latter could play a role analogous to that of an Immirzi parameter \cite{Immirzi:1996di} in $(2+1)$ dimensions \cite{Meusburger:2008dc,Bonzom:2008tq}.  In other words the following general action \cite{Solodukhin:2005ah,Miskovic:2006tm}   for the intermediate values of the standard/exotic BTZ black hole would be a linear combination of $I_{EH}$ and $I_{GCS}$
\begin{eqnarray}
    I(A)=& & \frac{\alpha}{8\pi G_N} 
    \left[
    \int_{M} \left(
     e^{a} \wedge R_{a} + \frac{\Lambda}{6}\epsilon_{abc} e^{a}\wedge e^{b} \wedge e^{c}
    \right)  +   {\frac{1}{2} \int_{\partial M} K } \right] \nonumber \\
    & & + \frac{\gamma \ell}{8\pi G_N}  \int_{M} \left(
    \omega^{a} \wedge d\omega_{a} + 
 {\frac{1}{3}} 
    \epsilon_{abc} \omega^{a}\wedge \omega^{b} \wedge \omega^{c} + \Lambda e^{a}\wedge T_{a}
    \right)
 \label{genact}   
\end{eqnarray}
 where $K$ refers to the trace of the second fundamental form at the boundary $\partial M$
 and was shown to regularize $(2+1)$ dimensional gravity \cite{Banados:1998ys}.
 The contribution to the WdW patch would therefore be the sum of these   contributions (with the proper normalization of the   {dimensionless} $\alpha$ and $\gamma$ coefficients).

The sum can be obtained by recalling that  a $(2+1)$ dimensional space of constant negative curvature has  symmetry
 $SO(2, 2)$, for which the generators of the Lie group satisfy \cite{Witten:1998zw}
\begin{eqnarray}
    \label{eq:CMR}
    \left[J_{a},J_{b} \right]=\epsilon_{abc} J^{c},\qquad 
    \left[J_{a}, P_{b} \right]=\epsilon_{abc}P^{c},\qquad
    \left[ P_{a}, P_{b} \right]=\frac{1}{\ell^2} \epsilon_{abc} J_{c}
\end{eqnarray}
with general invariant quadratic forms
\begin{eqnarray} \label{bilin}
    \langle{J_{a},J_{b}} \rangle = \gamma \eta_{a b}, \qquad
    \langle{J_{a},P_{b}}\rangle = \alpha \eta_{ab}, \qquad
    \langle{P_{a},P_{b}} \rangle = \gamma \eta_{ab}
\end{eqnarray}
where the constants $\alpha$ and $\gamma$ are introduced in Eq. \eqref{genact} and should be fixed to zero respectively if one wants to separately analyze only the Einstein-Hilbert case (EH) or the gravitational Chern Simon (GCS) case. 
 
The shift \eqref{SystM}, \eqref{SystJ} was   found in \cite{Kraus:2005zm,Solodukhin:2005ah} and   is given by the fact that the action \eqref{genact} describes a class of theories with topological mass, where the mass and  angular momentum of the BTZ black hole are linear combinations of the mass and angular momentum that can be calculated in pure GR.
 The variation of the total action \eqref{genact} with respect to the metric gives the following equations
\be
    R_{\mu \nu} - \frac{1}{2} g_{\mu \nu} R -g_{\mu \nu} \ell +\gamma C_{\mu \nu} = 0
\ee
where $C_{\mu \nu}$ is the Cotton tensor obtained varying the gravitational Chern-Simons action with respect to the metric.

\subsection{Thermodynamics}

{A full treatment of the extended phase space would entail a separate thermodynamic interpretation
of  $ \tilde{\ell}$, $\hat{\ell}$, and  $\ell$ as distinct thermodynamic quantities, each with their own conjugate.  The net effect would be to introduce three distinct pressures, each with their own conjugate volume.  This will add unnecessary complications to the basic thermodynamics of the generalized exotic BTZ black hole, and so for simplicity we shall consider $ \tilde{\ell} =\hat{\ell} = \ell$ henceforth.}

The Hawking temperature and the angular velocity associated with the event horizon read
\begin{eqnarray}
    T = \frac{r_{+}}{2 \pi  \ell^2}-\frac{8 G_N^2 j^2}{\pi  r_{+}^3}\,,\;\; \label{temp} \;\;
    \Omega = 4\frac{G_N j}{ r_{+}^2}\,, \label{thermovars1}
\end{eqnarray}
and are obtained from \eqref{BTZ1} using the horizon equation $N=0$.
These expressions for these intensive thermodynamic variables are geometric and model independent \cite{Townsend:2013ela}, {\it i.e.} independent of the particular field equations that are solved by the BTZ metric.  They only depend on the position of the Killing horizon.

\begin{figure}[h]
\begin{center}
\includegraphics[height=7cm]{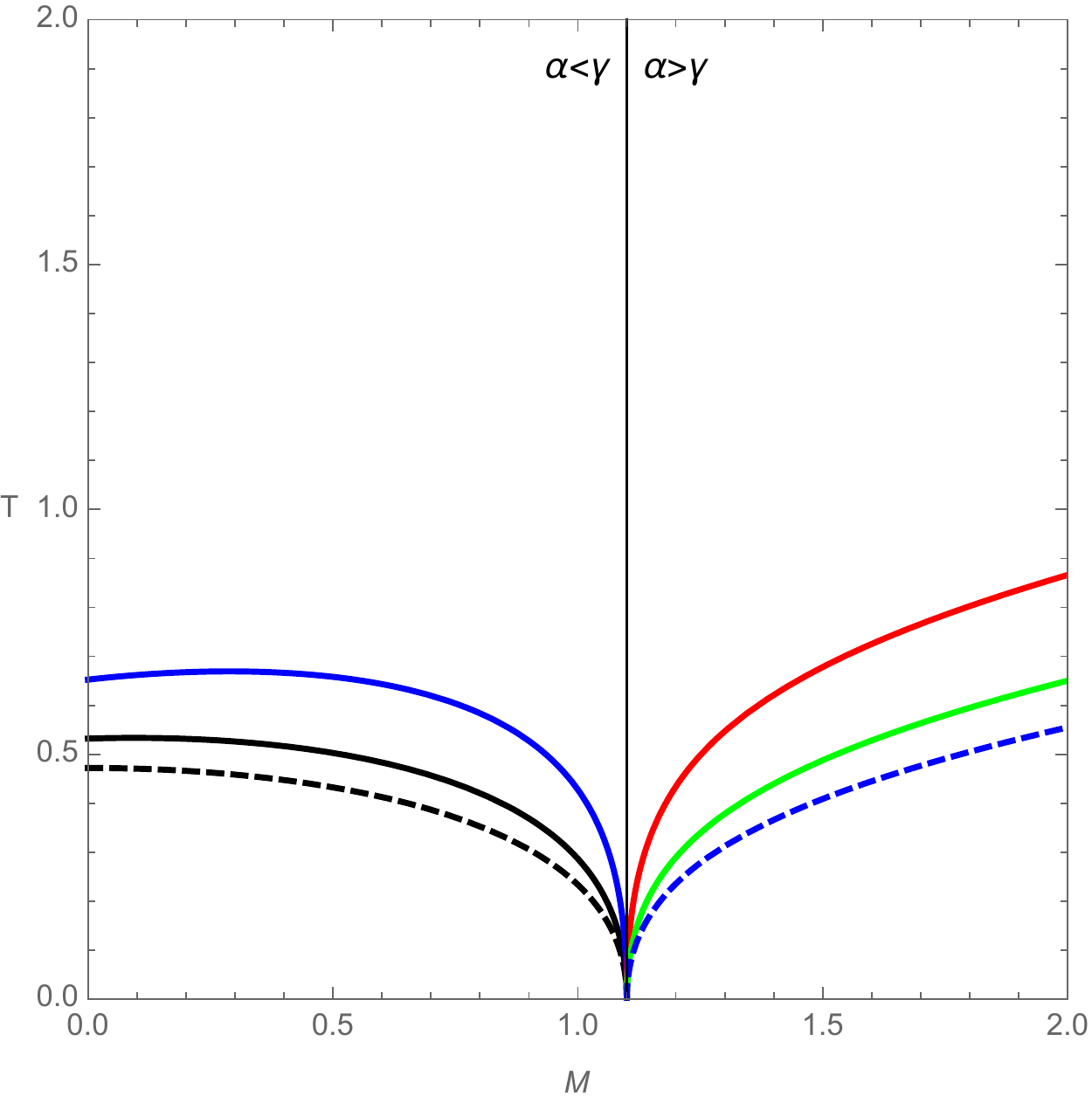}
\includegraphics[height=6.9cm]{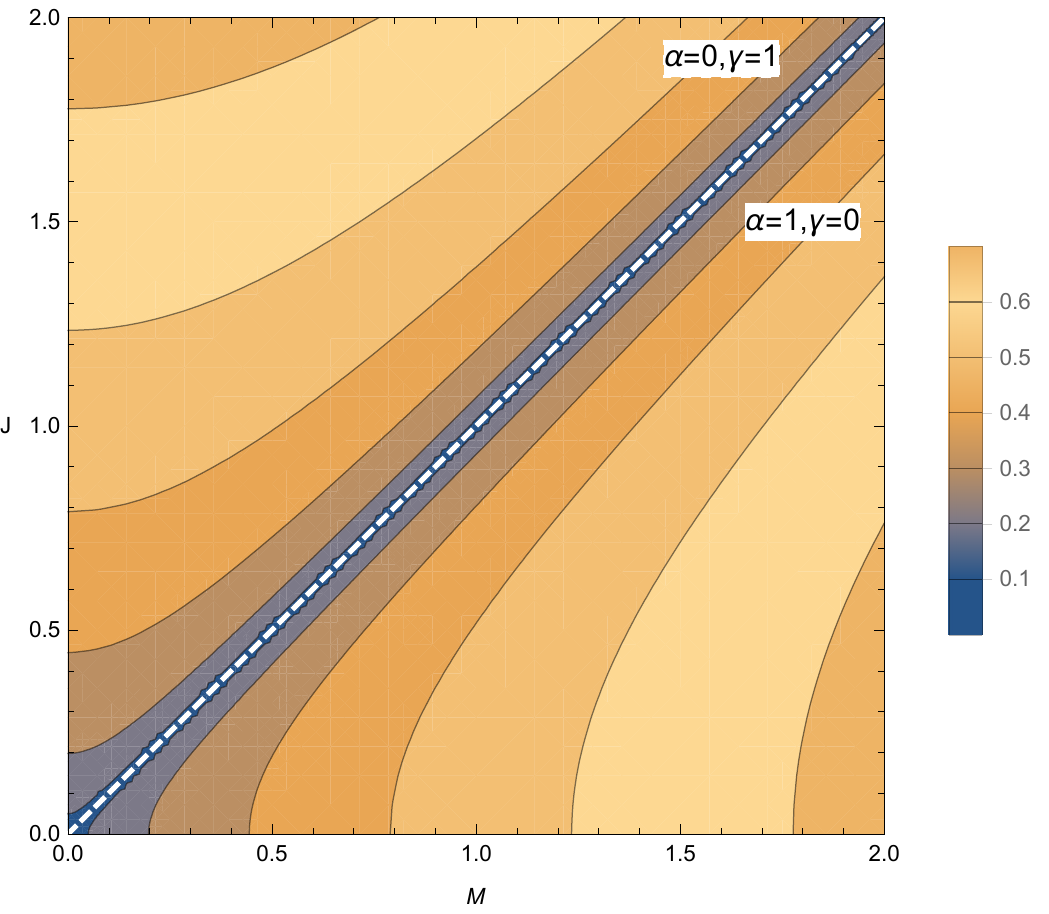}
\caption{
\emph{Left:} The two different sectors of the $(T,M)$ parameter space are divided by the vertical black line, which corresponds to the extremal case $\ell M = J$. The area on the left ($\alpha<\gamma$)   admits the exotic solution $\gamma=1,\alpha=0$ (black dashed line) for $J=1.1$ whereas the area on the right ($\alpha>\gamma$)   admits the standard BTZ solution $\gamma=0,\alpha=1$ (blue dashed line) for $J=1.1$.  The black/blue solid lines at the left are the respective cases ($\alpha=0.2,\gamma=0.6$) and ($\alpha=0.4,\gamma=0.6$), whereas the red and green solid lines respectively correspond to ($\alpha=0.6,\gamma=0.4$) and ($\alpha=0.8,\gamma=0.2$).
\emph{Right:} Contour plot of the temperature as function of $J$ and $M$ for the two sectors: the exotic and the standard case. 
  }
\label{Fig:TvsM}
\end{center}
\end{figure}

From \eqref{SystM} and \eqref{SystJ} we see that the extensive thermodynamic variables are model-dependent.   
Furthermore, the system of equations \eqref{SystM}, \eqref{SystJ} is singular when $\alpha=\gamma$. However, in the particular case $\ell m = j$ (that is the extremal solution) the condition $\ell M = J$ also holds for every value of $\alpha$ and $\gamma$ (including $\alpha=\gamma$). Likewise  if $\alpha=\gamma$ then $\ell M = J$, regardless of the distinct values of $m$ and $j$.
Rewriting the condition $\ell m \geq j$ in terms of the new variables $(M,J)$ we obtain two inequalities
\be\begin{cases} \label{eq:cases}
\ell M \geq J, & \alpha \geq \gamma \\
\ell M \leq J, & \alpha \leq \gamma 
\end{cases}
\ee
and   the condition needed to find the horizon is different for each. This   general analysis is valid for the redefinition of the conserved charges in Eqs. \eqref{SystM}, \eqref{SystJ} and  it is not necessary to have an extended phase space. In  Figure~\ref{Fig:TvsM} we show the behaviour of the temperature for the two different situations.

\subsection{Extended Phase Space}

Based on the hypothesis \eqref{pressD}, we now add to the intensive variables \eqref{thermovars1} another intensive quantity 
\begin{equation}
    P = \frac{1}{8 \pi G_N \ell^2}
    \label{thermovars2}
\end{equation}
 that is only geometric  and model independent. The first law of black hole thermodynamics \eqref{TDD} is 
\begin{equation}
    dM= TdS + VdP + \Omega dJ \label{eq:1law}
\end{equation}
and we wish to check this law for general values of $(\alpha,\gamma)$.

Inserting \eqref{mANDj} in the previous general definitions \eqref{SystM}, \eqref{SystJ}  of $(M,J)$ that interpolate between the classical and the exotic BTZ black hole we obtain
\begin{eqnarray}
     M &=& \frac{\alpha  \left(r_{-}^2+r_{+}^2\right)}{8 G_N \ell^2}+\frac{\gamma  r_{-} r_{+}}{4 G_N \ell^2}\,, \label{Mdef}\\
     J &=&  \frac{\alpha  r_{-} r_{+}}{4 G_N \ell}+\frac{\gamma  \left(r_{-}^2+r_{+}^2\right)}{8 G_N \ell} \label{Jdef} \\
     T &=& \frac{r_{+}^2-r_{-}^2}{2 \pi  \ell^2 r_+}   \qquad  \Omega  = \frac{r_-}{r_+ \ell}  \label{TOdef}
\end{eqnarray}
Since $M$ is a function of $\left( r_{+}, r_{-}, \ell \right)$ then
\begin{eqnarray}
    dM  &(&r_{+},r_{-},\ell) = \left( \frac{\partial M}{\partial r_-} \right) dr_{-}+ \left( \frac{\partial M}{\partial r_+} \right) dr_{+}+ \left( \frac{\partial M}{\partial \ell} \right) d \ell =\\
    &=& \left( \frac{\alpha  r_{-}}{4 G_N \ell^2}+\frac{\gamma  r_{+}}{4 G_N \ell^2} \right) dr_{-} +
        \left( \frac{\alpha  r_{+}}{4 G_N \ell^2}+\frac{\gamma  r_{-}}{4 G_N \ell^2} \right) dr_{+} + \left( -\frac{\alpha  \left(r_{-}^2+r_{+}^2\right)}{4 G_N \ell^3}-\frac{\gamma  r_{-} r_{+}}{2 G_N \ell^3} \right) d \ell \nonumber \\
    &=& \frac{1}{4 G_N \ell^2} \left[\left( {\alpha  r_{-}}+{\gamma  r_{+}} \right) dr_{-} +
        \left( {\alpha  r_{+}}+{\gamma  r_{-}} \right) dr_{+} \right] +
        \left( -\frac{\alpha  \left(r_{-}^2+r_{+}^2\right)}{4 G_N \ell^3}-\frac{\gamma  r_{-} r_{+}}{2 G_N \ell^3} \right) d \ell   
 \nn        
\end{eqnarray}
Expanding the right-hand-side of \eqref{eq:1law}, we have
\begin{eqnarray}
    TdS &=& \frac{r_{+}^2-r_{-}^2}{2 \pi  \ell^2 r_+} \left[
    \left( \frac{\partial S}{\partial r_{-}} \right) d r_{-} +
    \left( \frac{\partial S}{\partial r_{+}} \right) d r_{+} 
      \right]\\
     \Omega dJ &=&  \frac{\Omega}{4G_N\ell} \left[\left( \gamma r_{-} + \alpha r_{+}  \right) dr_{-} + 
     \left( \alpha r_{-} + \gamma r_{+} \right) dr_{+}  
     - \frac{ \gamma r_{-}^{2} + 2 \alpha r_{-} r_{+} + \gamma r_{+}^{2}}{2\ell}  d\ell
     \right]\\
     VdP &=& V  \left(-\frac{1}{4 \pi G_N \ell^3}\right) d\ell
\end{eqnarray}
where $\Omega = r_{-}/ \left( \ell r_{+}\right)$ and we suppose that the entropy has no dependence from the cosmological constant $S(r_{+},r_{-})$. 

Equating the terms for $dr_{+}$,  $dr_{-}$ and $d \ell$ in \eqref{eq:1law}, yields
\be 
    \frac{\partial S}{\partial r_{+}} = \frac{\pi  \alpha }{2G_N}\qquad
    \frac{\partial S}{\partial r_{-}} =  \frac{\pi  \gamma}{2G_N} \label{entropies} 
\ee    
and so
\begin{eqnarray}
\label{entropy1}
S &=&\frac{1}{2 G_N} \left(  \pi  \alpha  r_++\pi  \gamma  r_- \right) \\
V &=&     \alpha \pi r_+^2
     +  \gamma \pi r_{-}^2   \left( \frac{3  r_{+}}{2 r_{-}} - \frac{ r_{-}}{2 r_{+}}
      \right) \label{eq:vf}    
\end{eqnarray}
for the entropy and volume respectively.   For the standard BTZ black hole $\left( \gamma = 0, \alpha =1 \right)$ we obtain $V = \pi  r_+^2 $ (which happens to be the geometric volume of the black hole), while in the exotic case $\left( \gamma = 1, \alpha =0 \right)$ we get a mixed term that depends on both the inner and outer radii. The larger class of generalized exotic BTZ black holes therefore furnish  another example of thermodynamic volume that is different from the standard geometric volume.  
Note that in the extremal case ($r_{+}=r_{-}=r_{ext}$) the volume is $V=\pi r^2_{ext} (\alpha + \gamma)$. It coincides with the geometric volume   for $(\alpha + \gamma) = 1$, which includes both the standard 
$(\alpha=1, \gamma=0)$ and exotic $(\alpha=0,\gamma=1)$ cases.

Using the above relations it is straightforward to verify that the  Smarr relation in $D=3$ dimensions
\begin{equation}\label{Smarr3d}
    0= TS - 2 PV + \Omega J
\end{equation}
also holds. We pause to comment that it is possible to obtain a general relationship between these quantities that \be\label{nonSmarr}
xM=TS + y\Omega J + zPV
\ee
where $\{x,y,z\}$ are constants that obey
\be
x-z = 2 \qquad  2y - z = 4
\ee 
 for all $(\alpha,\gamma)$.    More general relations of the type in equation \eqref{nonSmarr} are sometimes referred to as  Smarr relations \cite{Liang:2017kht,Erices:2017nta}, but they do not respect the correct Euler scaling relations unless $x=0$, forcing $y=1$ and $z=-2$, yielding
\eqref{Smarr3d}.

At this point, we can make two interesting considerations:
\begin{itemize}
    \item We required the validity of the laws \eqref{eq:1law}  and \eqref{Smarr3d} of  standard black hole thermodynamics to also hold in the case of the exotic black hole. This requirement gave us the correct entropy for the exotic BTZ black hole \cite{Carlip:1994hq,Townsend:2013ela,Solodukhin:2005ah}: 
    \begin{equation}
        I_{E}=\frac{\pi r_{-}}{2 G_N};
    \end{equation}
    \item The thermodynamic volume depends on both  variables $(\alpha, \gamma)$; in the exotic case it becomes
\begin{equation} \label{eq:vol}
      V =      \pi r_{-}^2  \left( \frac{3  r_{+}}{2 r_{-}} - \frac{ r_{-}}{2 r_{+}}
      \right).
\end{equation}
The volume is positive when the following condition is satisfied 
\begin{equation}
    r_{+}^2 \left(2 \alpha  r_{+} + 3 \gamma  r_{-}\right)>\gamma  r_{-}^3
\end{equation}
that in the case $\alpha=0$ becomes $r_{+}>r_{-}/\sqrt{3}$. It is always satisfied for the standard BTZ black hole, for which $\gamma=0$. 
\end{itemize}
So far our analysis holds for general values of  $(\alpha , \gamma)$.  For the remainder of this paper we shall assume normalization of the parameters such that $\alpha + \gamma = 1$.

\section{Pressure-Volume Behaviour}

It is possible to solve  \eqref{Jdef} and \eqref{eq:vf} 
 for both $r_+$ and $r_-$ in terms of $(P,V,J)$.  Inserting the results into \eqref{TOdef} we can solve for $P$ in
 terms of $(T,V,J)$ and obtain an equation of state for the generalized exotic black hole.
 
 We can preliminarily explore the expected $PV$ diagram characteristics of these systems by considering the small and large $J$ limits of the pure exotic and normal black hole equations of state.   Ignoring constants for the time being, setting $(\alpha = 1, \gamma=0)$ the equation of state takes the simple dependence
 \beq
P \sim \frac{J^2}{V^2} + \frac{T}{\sqrt{V}}~~,
 \eeq
 which strictly decreases for increasing $V$.  Particular values of $J$ and $T$ have no effect on the general qualitative shape of the corresponding $PV$ diagram.   
 
In the pure exotic limit $(\alpha=0, \gamma = 1)$, we find a much more complicated expression that exhibits a maximum volume when $J \gg T$.  Although the full expression is not trivial to solve, we can expand $V$ as a series in $T$ to third order, which gives
\beq
V \sim \frac{\sqrt{2.51 G_N} J}{\sqrt{P}} + \frac{0.351 T\sqrt{J}}{(G_NP)^{3/4}} - \frac{0.0491 T^2}{G_N^2P^2}
\eeq
which reaches a maximum when
\beq
P \sim \frac{0.138T^{4/3}}{G_N^{5/3} J^{2/3}}
\eeq

In this case, it becomes possible for  the system to admit two pressures for the same thermodynamic volume.   For small $J$, one can verify the volume behaves as $V \sim P^{-2}$.  Also, it can be shown that the angular momentum $J$ has a lower bound, which generalizes to arbitrary $\gamma$ as
\begin{equation}
    J \geq \frac{1}{32} \sqrt{\frac{\pi}{2}} \sqrt{\frac{T^4 \gamma^2}{G_N^5 P^3}}~~.
\end{equation}

The full equation of state for general $\alpha$ and $\gamma$ is analytically cumbersome. Rather than give the explicit expressions, we  shall compute the  $PV$ diagrams   parametrically.  Specifically, we focus on the trends in the implicit plots as the black holes go from being ``normal'' ($\alpha =1$) to exotic ($\alpha=0$), as well as for increasing angular momentum $J$ and temperature $T$.  
\begin{figure}[h]
\begin{center}
\includegraphics[scale=0.5]{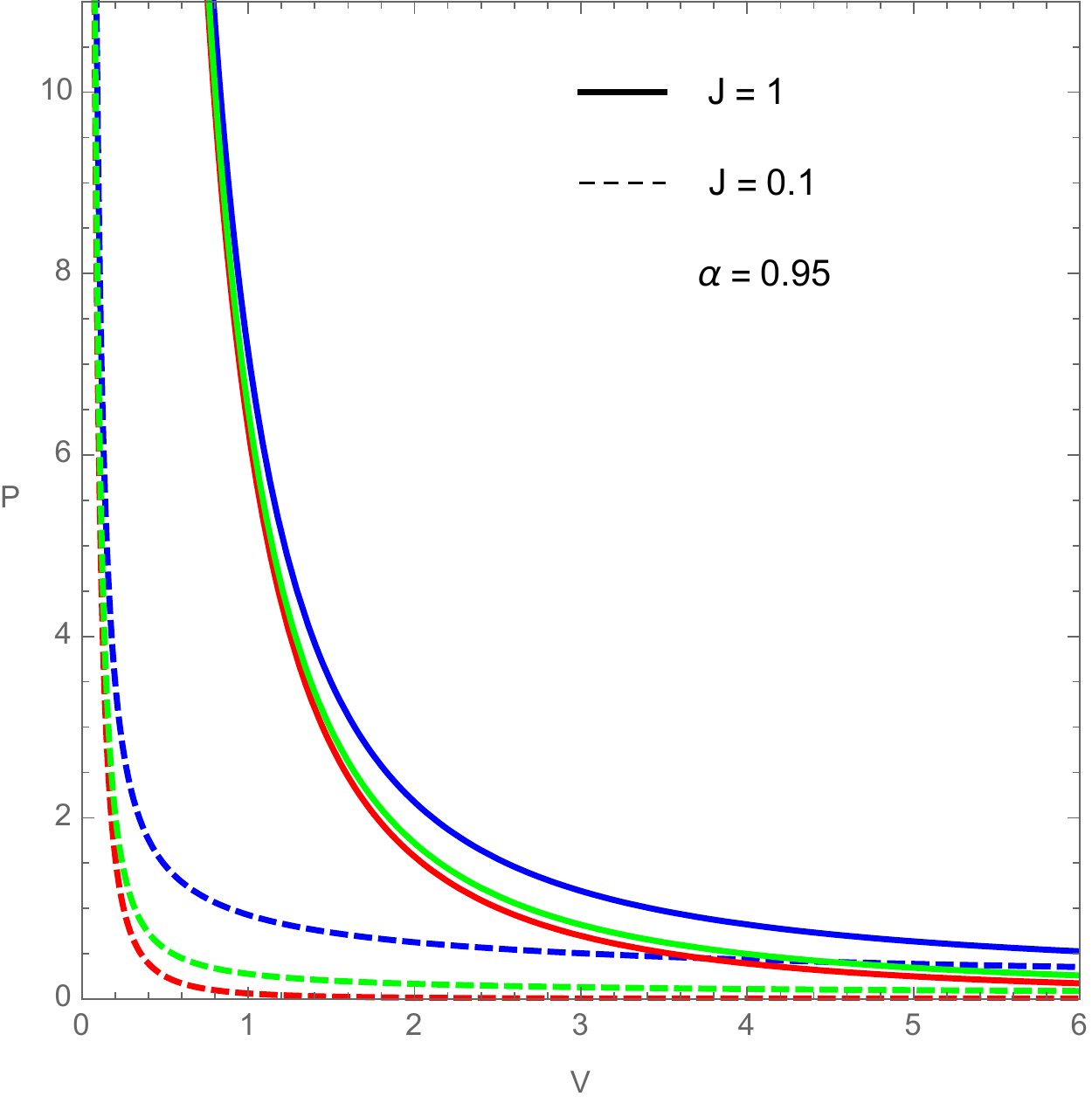} 
\includegraphics[scale=0.5]{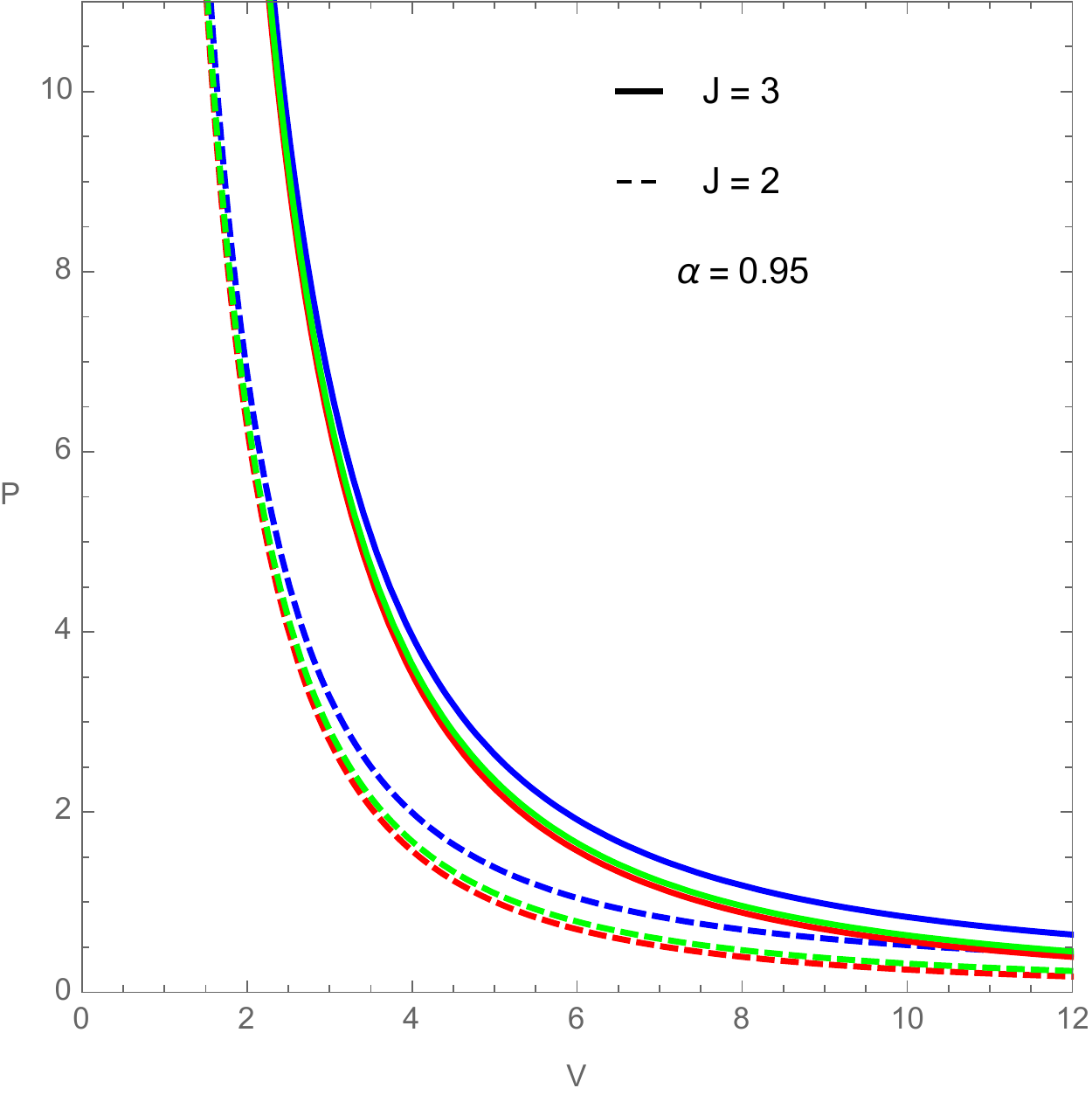} 
\caption{ PV diagram for $\alpha=0.95$, $\gamma=0.05$, $J = 0.1, 1$ (left)  and  $J=2, 3$ (right), and $T = 0.001$ (red), $T=0.5$ (green), $T=2$ (blue).}
\label{a90}
\end{center}
\end{figure}

\begin{figure}[h]
\begin{center}
\includegraphics[scale=0.5]{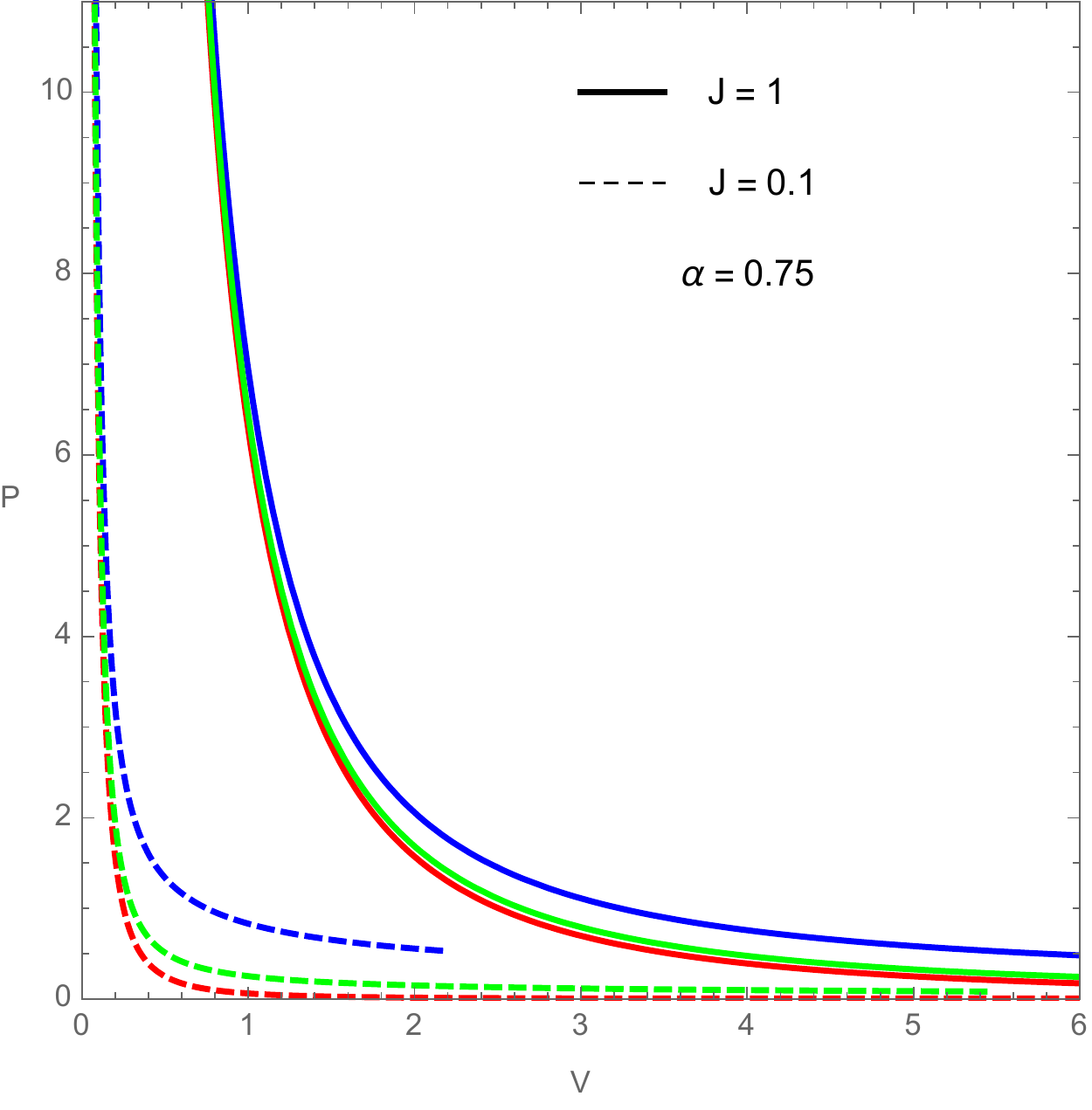} \includegraphics[scale=0.5]{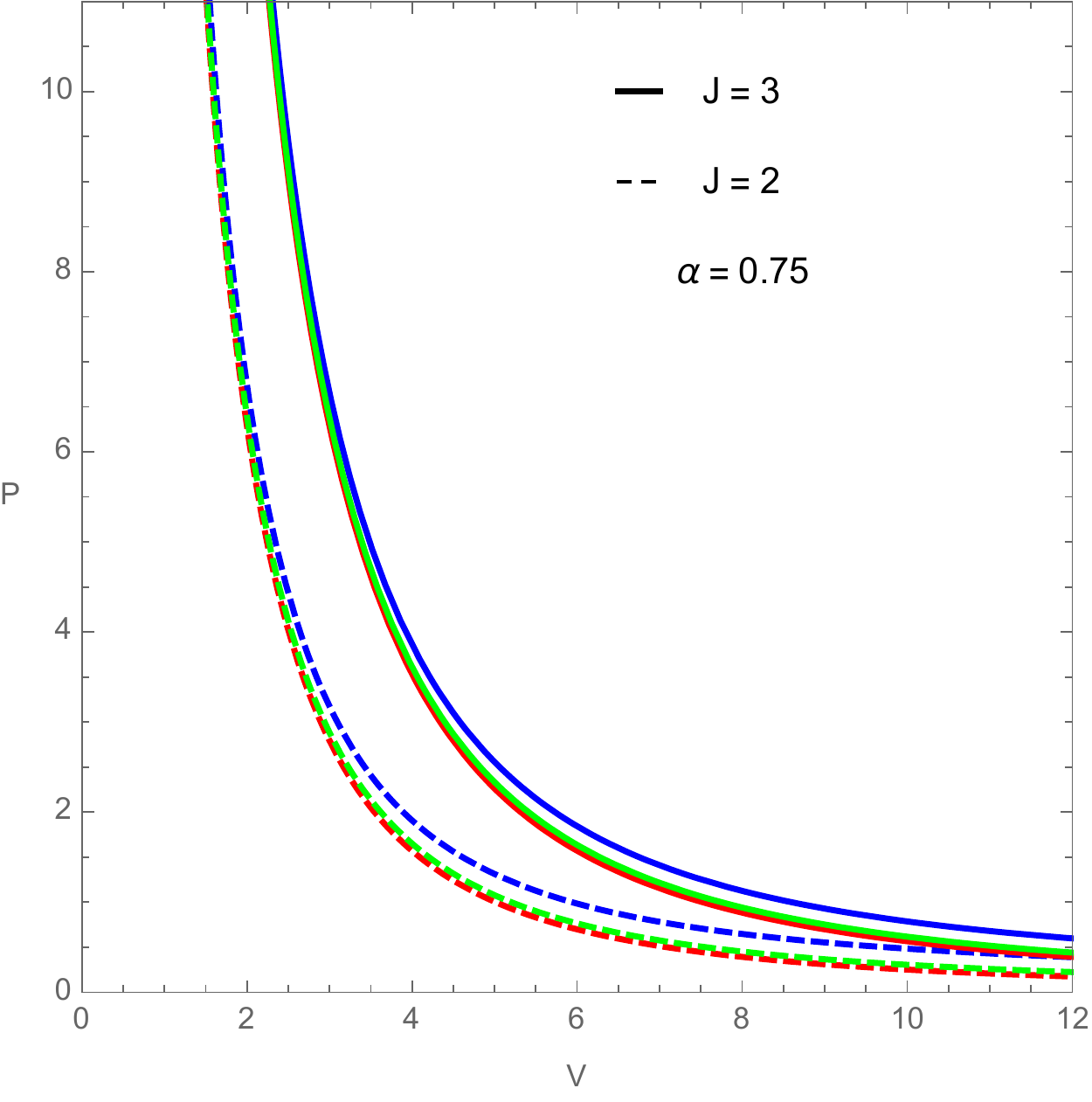}
\caption{PV diagram for $\alpha=0.75$, $\gamma=0.25$, $J = 0.1, 1, 2, 3$, and $T = 0.001$ (red), $T=0.5$ (green), $T=2$ (blue). In the first panel it is possible to see where the solution that satisfies Eq. \eqref{eq:cases} ceases to exist. }
\label{a75}
\end{center}
\end{figure}

 To construct the plots we solve the system of equations \eqref{Jdef} and \eqref{TOdef} for both $r_+$ and $r_-$, inserting the results in the definition of the thermodynamic volume \eqref{eq:vol} in order to obtain the $PV$ plot.
There are two different solutions depending on whether $\alpha>\gamma$ or $\alpha<\gamma$.
The cuts in the following plots are given by the above lower bound on $J$.

Figure~\ref{a90} illustrates the characteristics of a  95\% regular black hole   ($\alpha=0.95$, $\gamma=0.05$) 
for increasing $J$.  Each curve is isothermal, with the temperature increasing    from left to right.  There is no apparent critical behaviour, nor any change in the general shape of the curves, with only their concavity decreasing as $J$ grows.  Figure~\ref{a75} depicts this relation for the same parameters for a 75\% regular ($\alpha=0.75$)
black hole.  There is no significant deviation between the three cases.    
\begin{figure}[h]
\begin{center}
\includegraphics[scale=0.5]{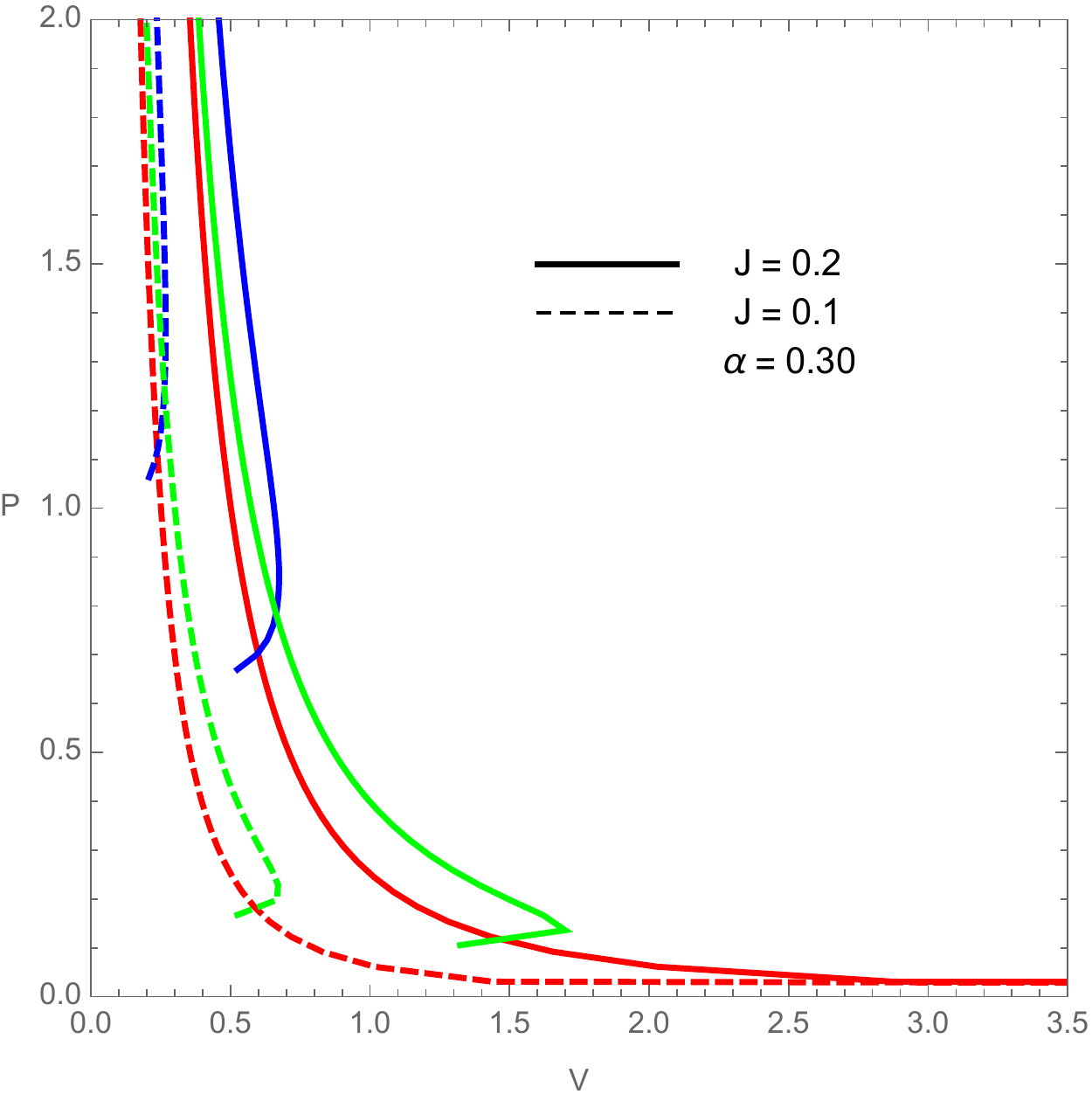} 
\includegraphics[scale=0.5]{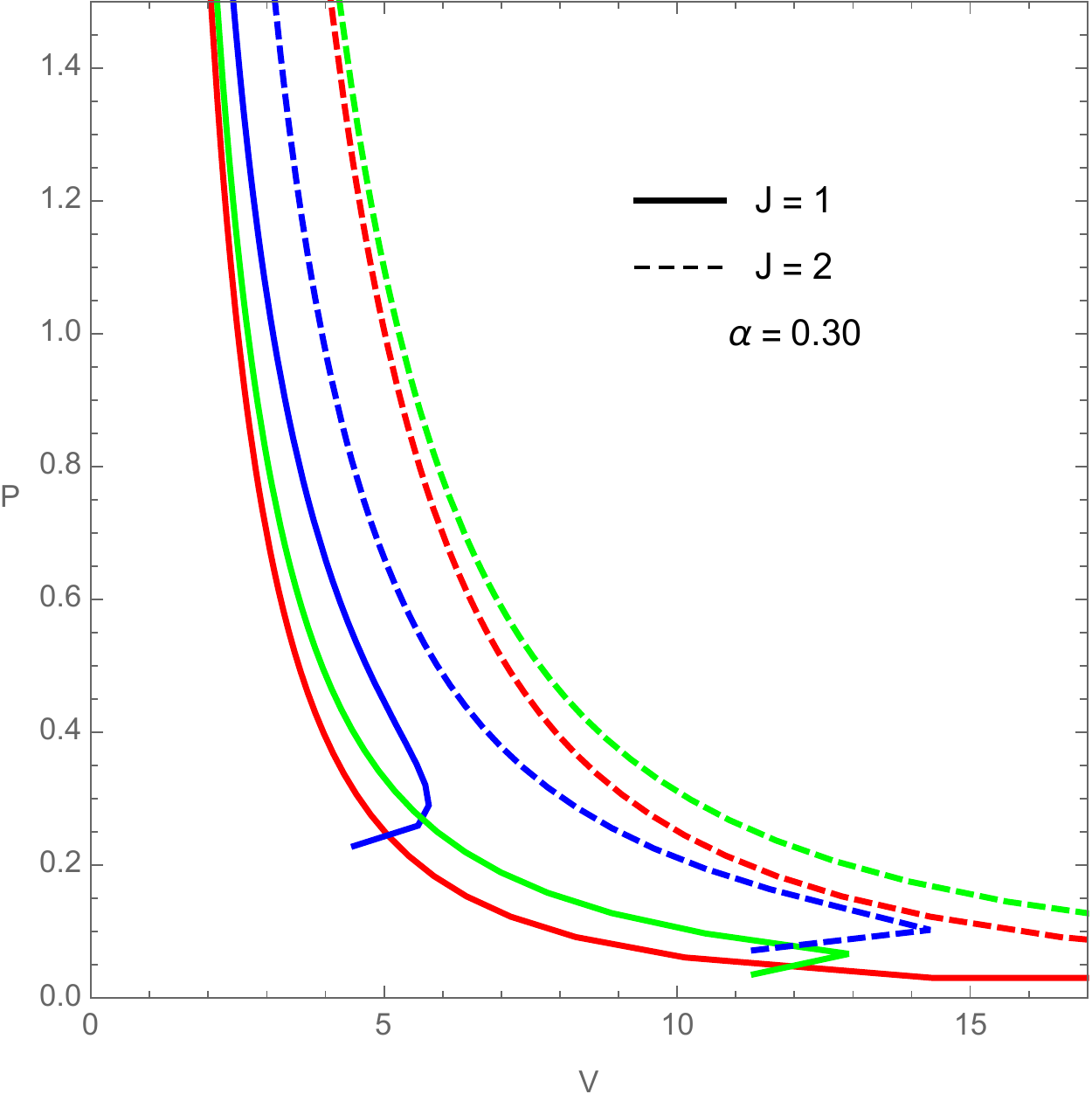}
\caption{PV diagram for $\alpha=0.30$, $\gamma=0.70$, $J = 0.1, 0.2, 1, 2$, and $T = 0.001$ (red), $T=0.5$ (green), $T=2$ (blue).}
\label{a30}
\end{center}
\end{figure}

\begin{figure}[h]
\begin{center}
\includegraphics[scale=0.5]{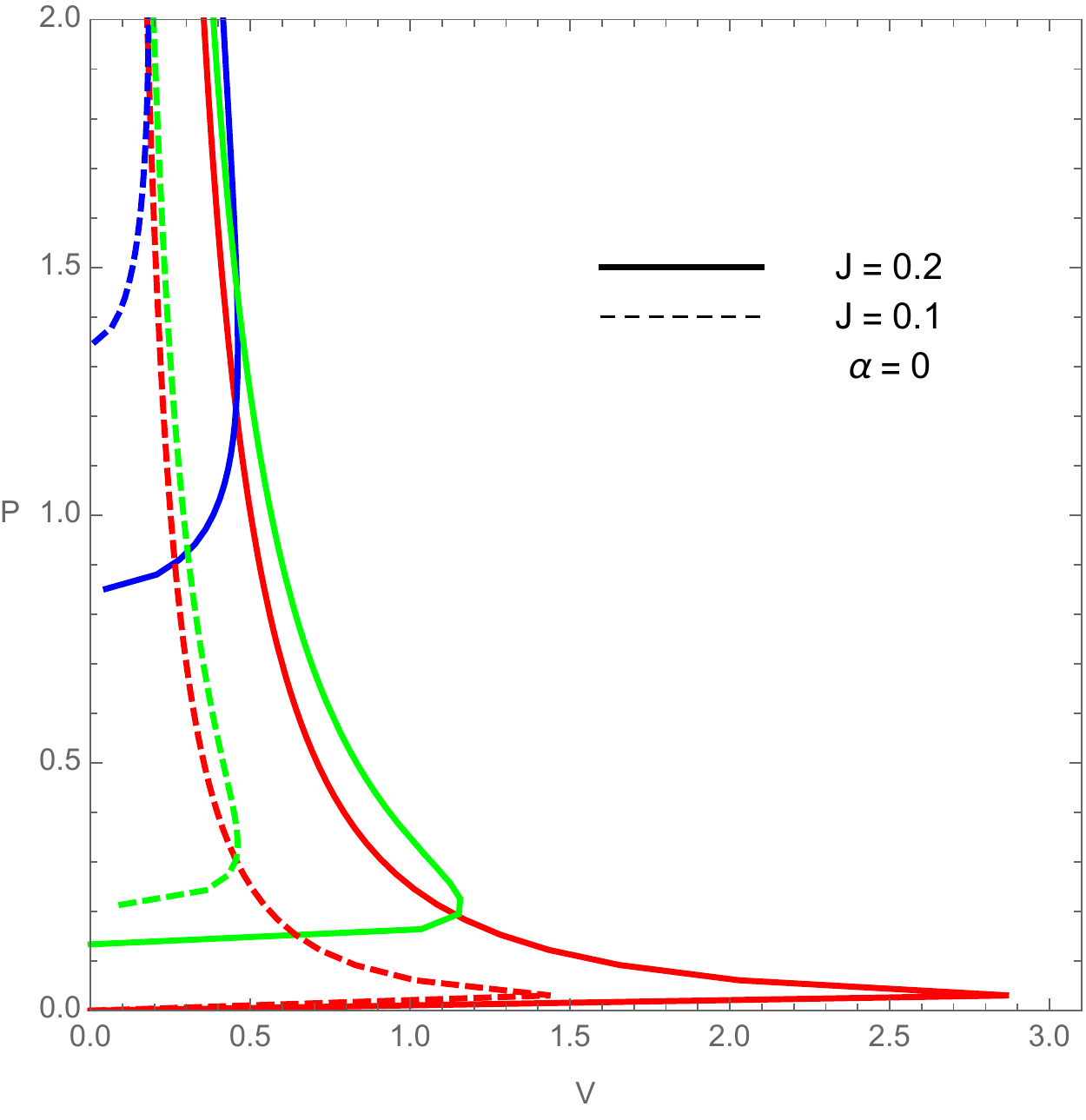} 
\includegraphics[scale=0.5]{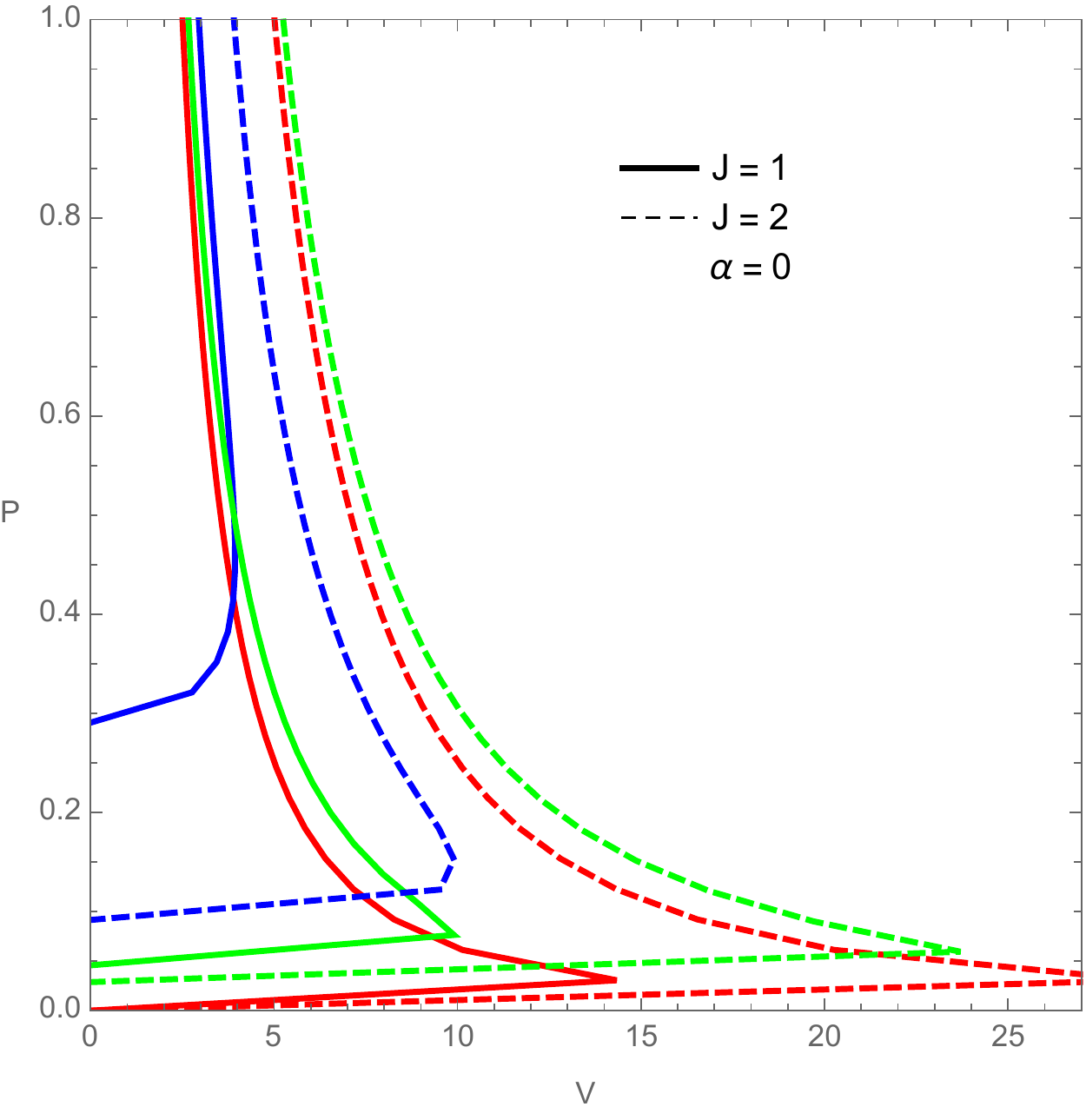}
\caption{PV diagram for $\alpha=0$, $\gamma=1$ (pure exotic BTZ black hole). The other parameters are $J = 0.1, 1, 2, 3$, and $T = 0.001$ (red), $T=0.5$ (green), $T=2$ (blue). In the case of exotic BTZ there is a value of the pressure for which the volume is equal to zero. }
\label{a00}
\end{center}
\end{figure}

 Interesting behaviour begins to emerge for smaller $\alpha$.  In  Figure~\ref{a30}, for which $\alpha=0.30$, we see that in the limit of small $V$, the isothermal curves become double-valued.  This actually happens for the higher-$\alpha$ black holes as well, though it is at a much smaller volume scale as compared to the more exotic black holes.  In fact, as $J$ increases the overlap region of the isothermal curves becomes increasingly large (see Figure~\ref{a00}).  
This suggests that the angular momentum of the black holes plays a much more significant role in the chemistry of   exotic black holes, as compared to the normal ones.

In the case where $\alpha=0$, {\it i.e.} a purely exotic BTZ black hole, the phase plots show similar behaviour as above, only more pronounced (see Figure~\ref{a00}).  For sufficiently large $T$ and small $J$, 
the  associated pressure is a smoothly increasing function of $V$.  As temperature decreases, a critical value is reached at which the  curve becomes double-valued -- for small enough $V$ two states can exist with different pressures.  For large enough $J$ and small enough $T$ this double-valuedness is persistent.  As $V\to 0$ the pressure remains finite.

\section{Reverse Isoperimetric Ratio}

One quantity of interest for black holes is the reverse isoperimetric inequality  \cite{Cvetic:2010jb}, which is the conjecture that
\be\label{ratio}
{\cal R}= \Bigl( \frac{(D-1){V}\,}{\omega_{D-2} } \Bigr )^{\frac{1}{D-1}}\,
  \Bigl(\frac{\omega_{D-2}}{A}\Bigr)^{\frac{1}{D-2}} \geq 1
  \ee
 for  any asymptotically AdS black hole, with $A$ identified with the horizon area, $V$ with the associated thermodynamic volume, and $\omega_{D-2} = 2\pi^{\frac{D-1}{2}}/\Gamma\Bigl(\frac{D-1}{2}\Bigr)$.
   For Schwarzschild-AdS black holes
the bound is saturated, which means that for a fixed thermodynamic volume the
entropy of the black hole is maximized for   Schwarzschild-AdS spacetime.

\begin{figure}[h]
\begin{center}
\includegraphics[scale=0.3]{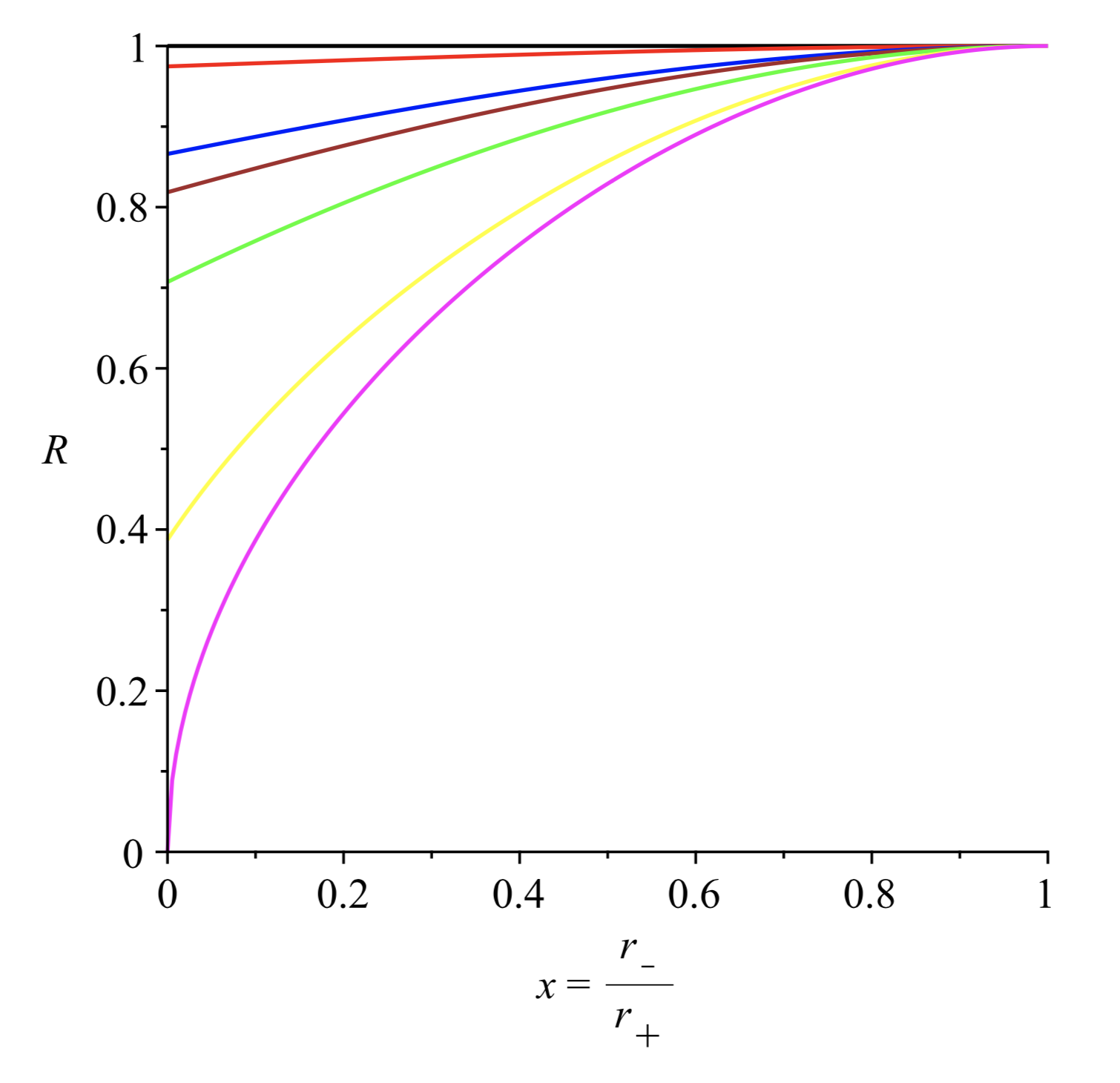}  
\caption{ Reverse isoperimetric ratio plotted as a function of  $x=r_-/r_+$ for various values of $(\alpha,\gamma)$:
$(1,0)$ (black), $(.95,.05)$ (red), $(.75,.25)$ (blue), $(.67,.33)$ (brown), $(0.5,0.5)$ (green)  $(.15,.85)$ (yellow), $(0,1)$ (magenta).
}
\label{Rir}
\end{center}
\end{figure}

For generalized exotic BTZ black holes we obtain 
\be
{\cal R} = {\frac{1}{2}}{\sqrt{\frac{4\alpha r_+^3 -2\gamma r_-^3 +6\gamma r_- r_+^2}{r_+^3}}}
\ee
which is saturated for the BTZ case $(\alpha=1,\gamma=0)$ but is otherwise violated for all 
other values of $(\alpha,\gamma)$. 
We illustrate this in Figure
\ref{Rir}.  

\section{Gibbs Free Energy and Critical Behaviour}

\begin{figure}[h]
\begin{center}
\includegraphics[scale=0.5]{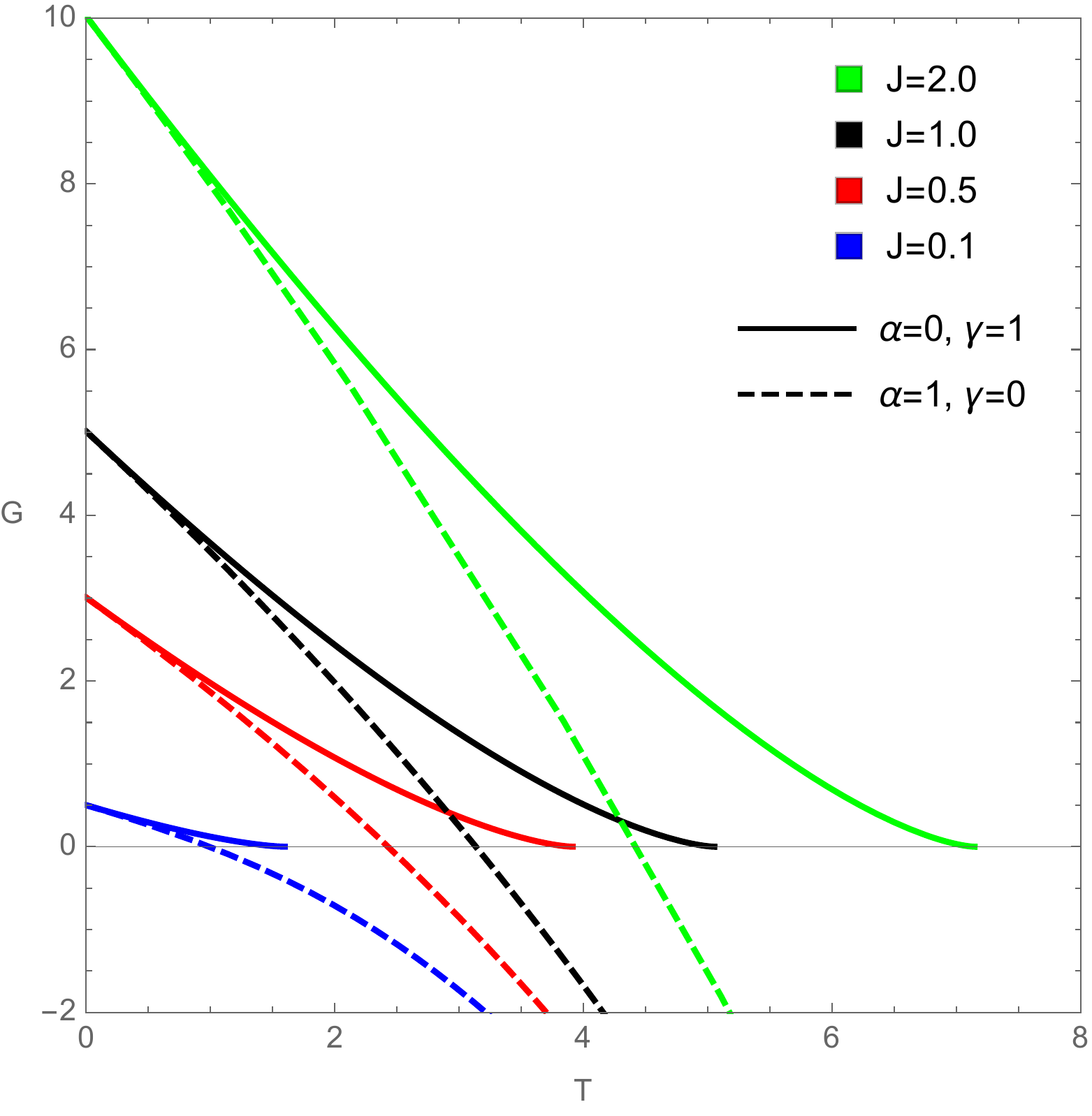}  
\caption{Gibbs free energy for a standard ($\alpha=1; \gamma=0$, dashed curve) and a purely exotic black hole ($\alpha=0; \gamma=1$, solid curve),
for angular momentum values $J=2.0$ (green), $1.0$ (black), $0.5$ (red), and $J=0.1$ (blue) and $P=1$.}
\label{gibbs1}
\end{center}
\end{figure}

\begin{figure}[h]
\begin{center}
\includegraphics[scale=0.5]{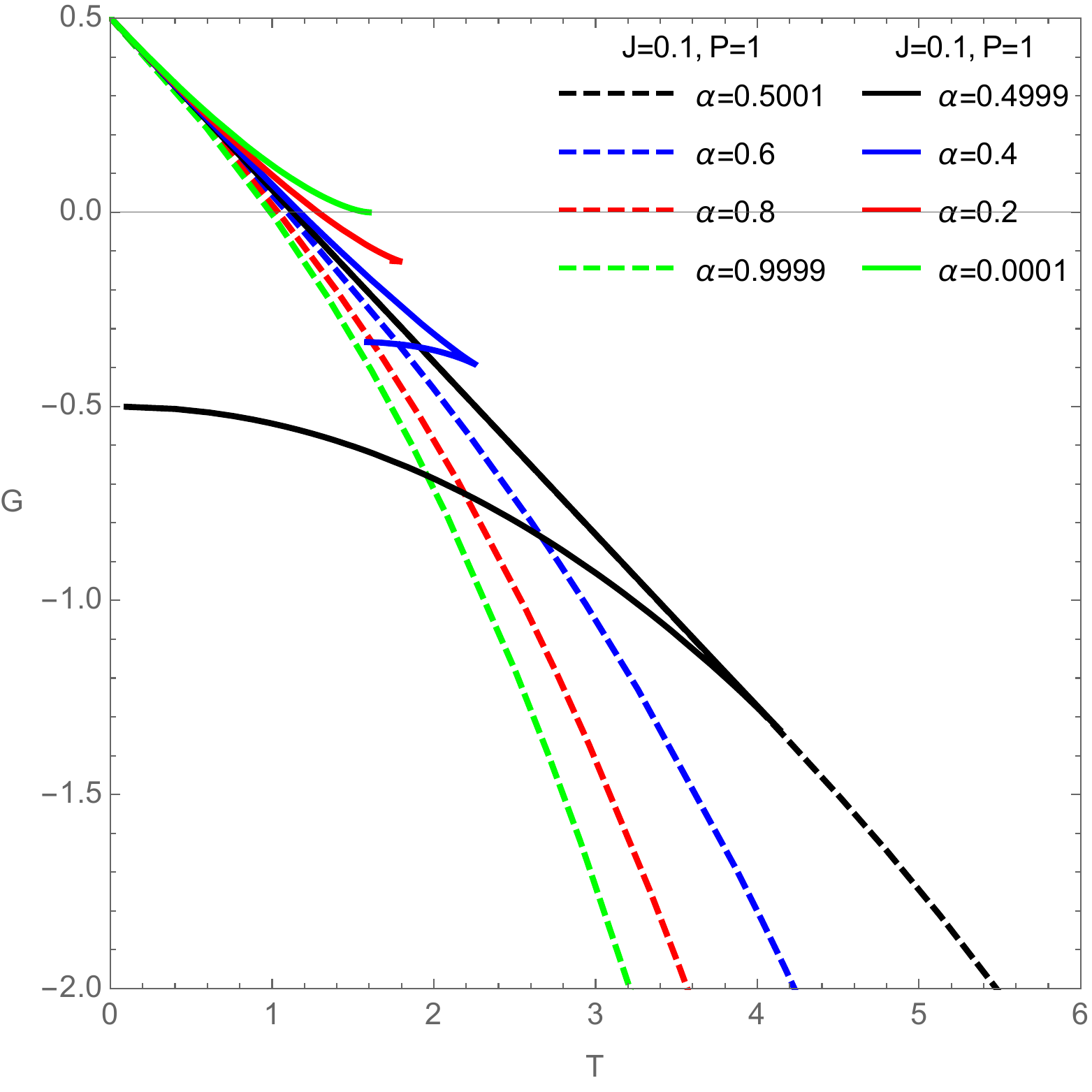}  
\caption{Gibbs free energy at the transition between majority standard ($\alpha > 0.5; \gamma < 0.5$) and majority exotic ($\alpha < 0.5; \gamma > 0.5$) black
holes.  }
\label{gibbs2}
\end{center}
\end{figure}

The Gibbs free energy of the black holes can be readily calculated, and the corresponding plots for different degrees of exoticness are displayed in Figures~\ref{gibbs1} and ~\ref{gibbs2}.  
The plots are obtained solving \eqref{Jdef} with respect to $r_{-}$ and substituting it in the temperature 
\eqref{TOdef} and in the Gibbs free energy 
\begin{equation}\label{Gibbs}
    G = M - TS = \frac{3 \alpha  r_{+} r_{-}^2-\alpha  r_{+}^3+2 \gamma  r_{-}^3}{8 G_N \ell^2 r_{+}}.
\end{equation}
There is a stark distinction between the cases where $\alpha > 0.5$ (majority standard) and $\alpha < 0.5$ (majority exotic).  For $\alpha < \gamma$, there is an upper bound for $r_{+}$ such that $r_{E} \leq r_{+} \leq r_{+_{M}}$  where    \begin{equation}
  r_{E} = {2 \sqrt{G_N J \ell}} \qquad    r_{+_{M}} 
     = 2 \sqrt{2} \sqrt{\frac{\gamma  G_{N} J \ell}{\gamma-\alpha}}
\end{equation}
are respectively obtained by   solving \eqref{Jdef} with respect to $r_{-}$ and  $T=0$ with respect to $r_{+}$ (so that $r_{E}$ is the same both in the standard and exotic cases) and using the relation $\left( \alpha + \gamma \right) =1$.

As shown in Figure~\ref{gibbs1}, the concavity of the two cases is opposite.  Moreover, the free energy of the purely exotic black hole  {terminates in $r_{+_{M}}$ } to zero for increasing temperature, while that of the standard black hole can be negative and is unbounded from below.

The intermediate cases where $0<\alpha,\gamma<1$ reveal even more interesting features for the majority exotic case.  As Figure~\ref{gibbs2} indicates, the energy curve reaches a minimum, but can assume two different values for the same temperature.  This behaviour is most pronounced for the transitional case at  $\alpha \rightarrow 0.5 $ with $\gamma = 1-\alpha$, and abates as $\alpha \rightarrow 0$. 
For $\alpha < \gamma$, the final points in Figures~\ref{gibbs1} and \ref{gibbs2} are  
\begin{eqnarray}
    T(r_{+_{M}}) &=&  
      \frac{\sqrt{2} \sqrt{\gamma  G_N J \ell \left(\gamma-\alpha\right)}}{\pi  \gamma ^2 \ell^2} 
    \\
    G(r_{+_{M}}) &=& -\frac{\alpha  J}{\gamma  \ell}
\end{eqnarray}
Local thermodynamic stability is given by  positivity of the specific heat, defined as
\begin{equation}
    C_P = T \left( \frac{\partial S}{\partial T}  \right)_{P,J}.
\end{equation}
Since the temperature has two different behaviours in the two different situations $\alpha>\gamma$  or $\gamma>\alpha$, we analyze the two cases separately.

\begin{figure}[h]
\begin{center}
\includegraphics[scale=0.55]{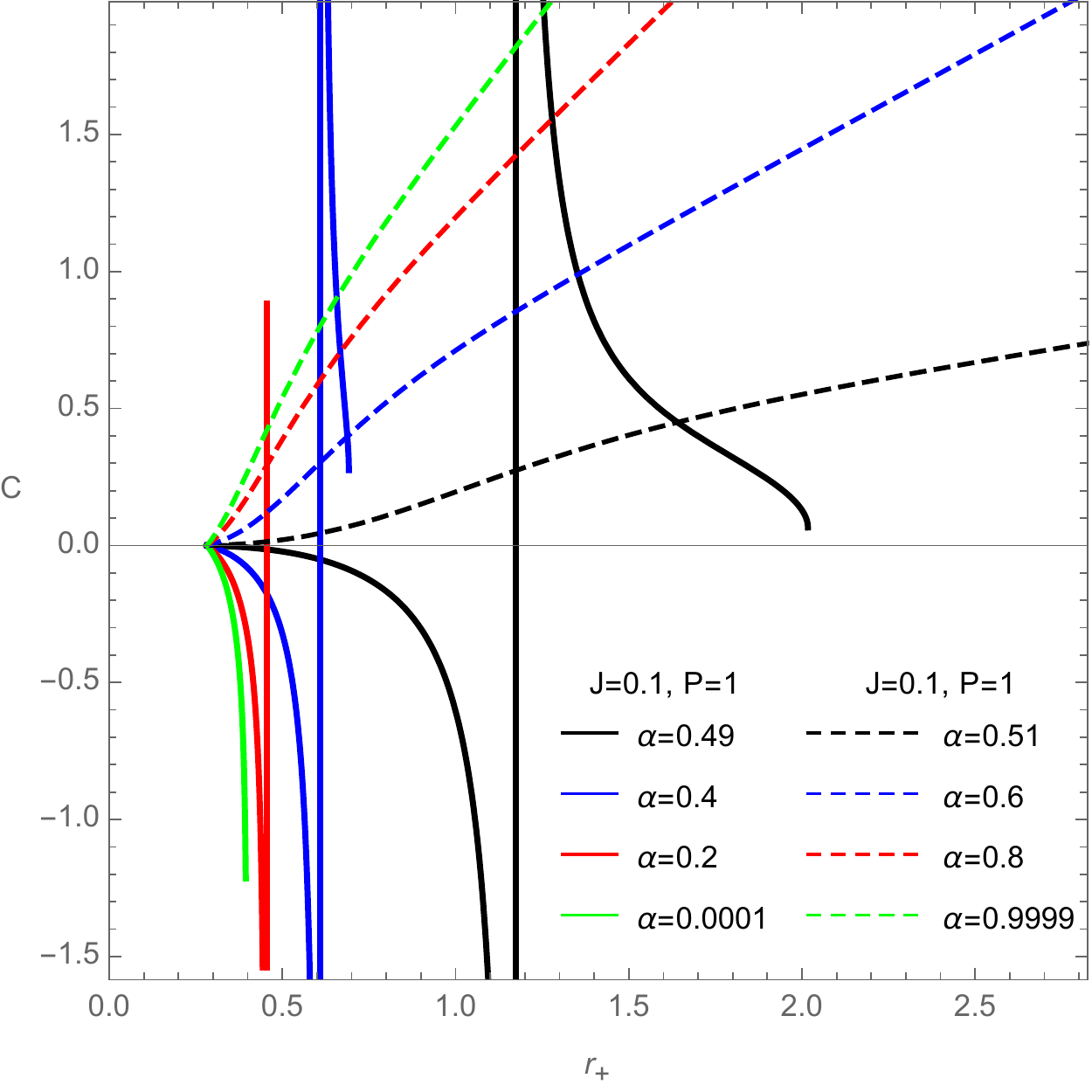} 
\includegraphics[scale=0.55]{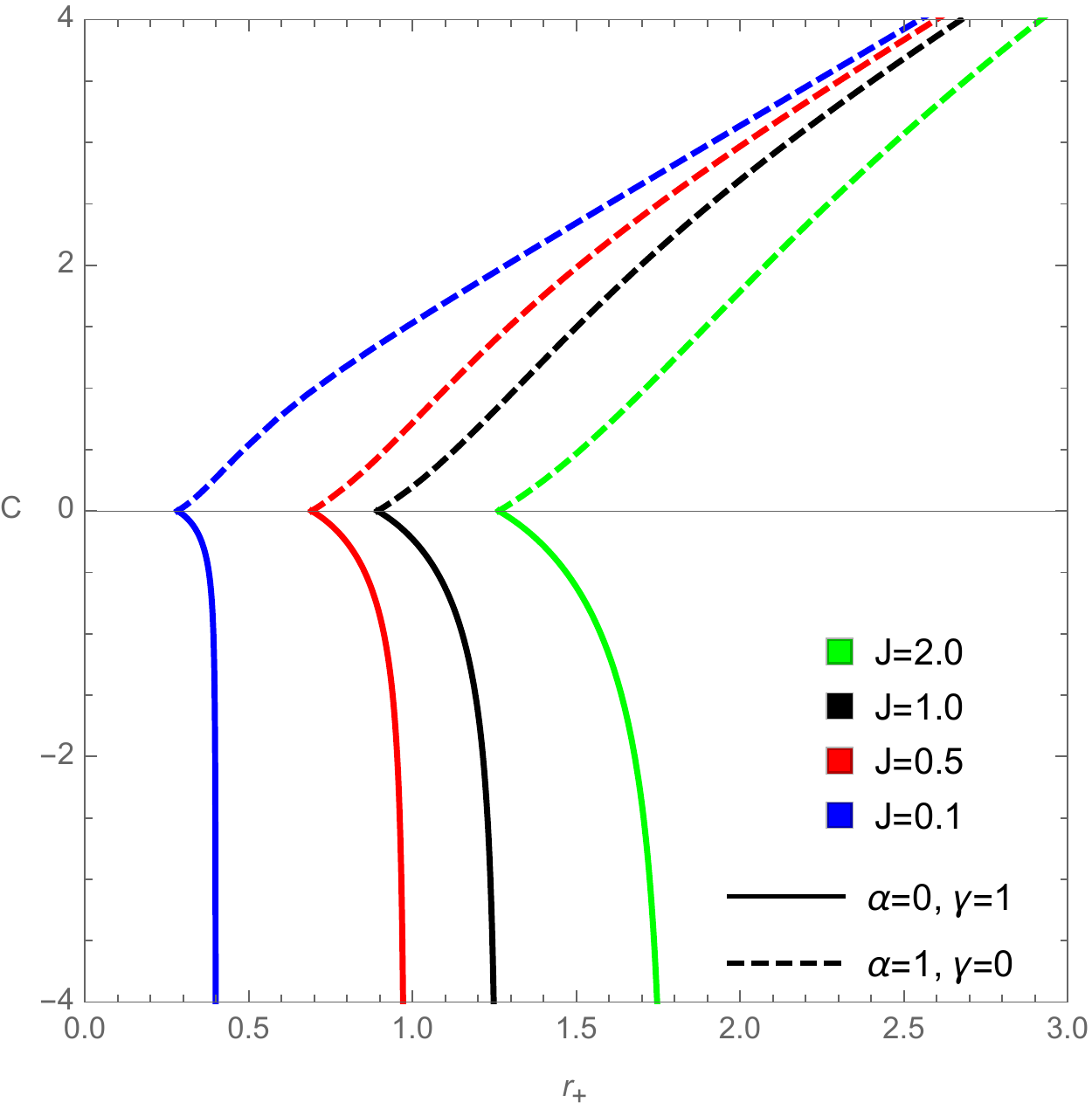}
\caption{    Specific heat: \emph{Left panel:} different $\alpha$'s and respective $\gamma=1-\alpha$. \emph{Right panel:} Pure exotic and standard case for several values of J.}
\label{Crp1}
\end{center}
\end{figure}

For $\alpha >\gamma$ and ($\alpha=1$, $\gamma = 0$) the specific heat is always positive (dashed lines in Figure~\ref{Crp1}).
The pure exotic case ($\alpha=0$, $\gamma=1$) always has negative specific heat therefore is locally unstable (see right panel in Figure~\ref{Crp1}). However, for intermediate values of $\alpha$ and $\gamma$ (with $\alpha<\gamma$) there are two branches (left panel in Figure~\ref{Crp1}), and the one with lower Gibbs free energy has positive specific heat while the other one has negative specific heat. The black holes with positive specific heat can be in stable equilibrium at a fixed temperature.  

\subsection{Free Energy with fixed-$\Omega$}
 
\begin{figure}[h]
\begin{center}
\includegraphics[scale=0.55]{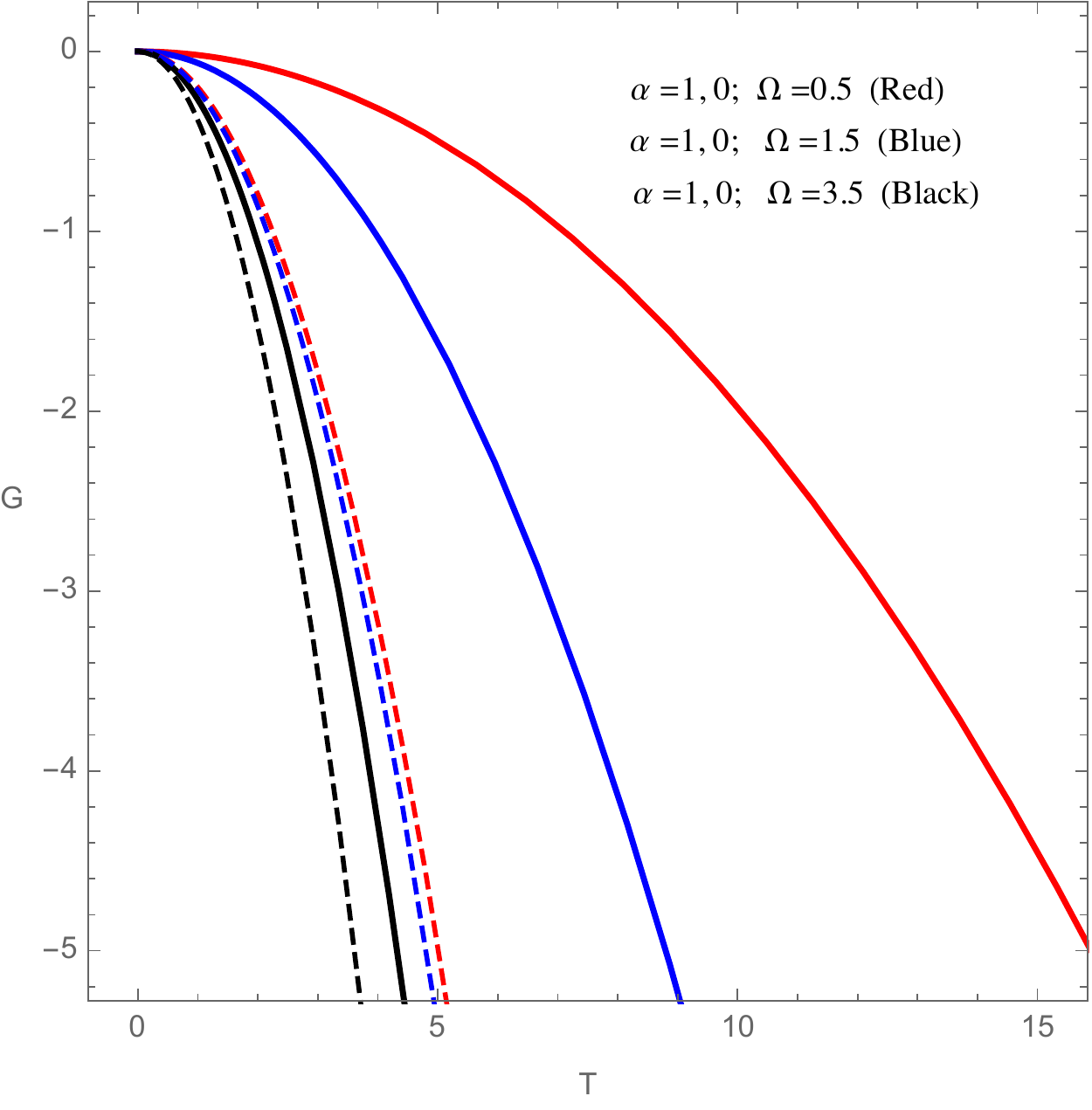} 
\includegraphics[scale=0.55]{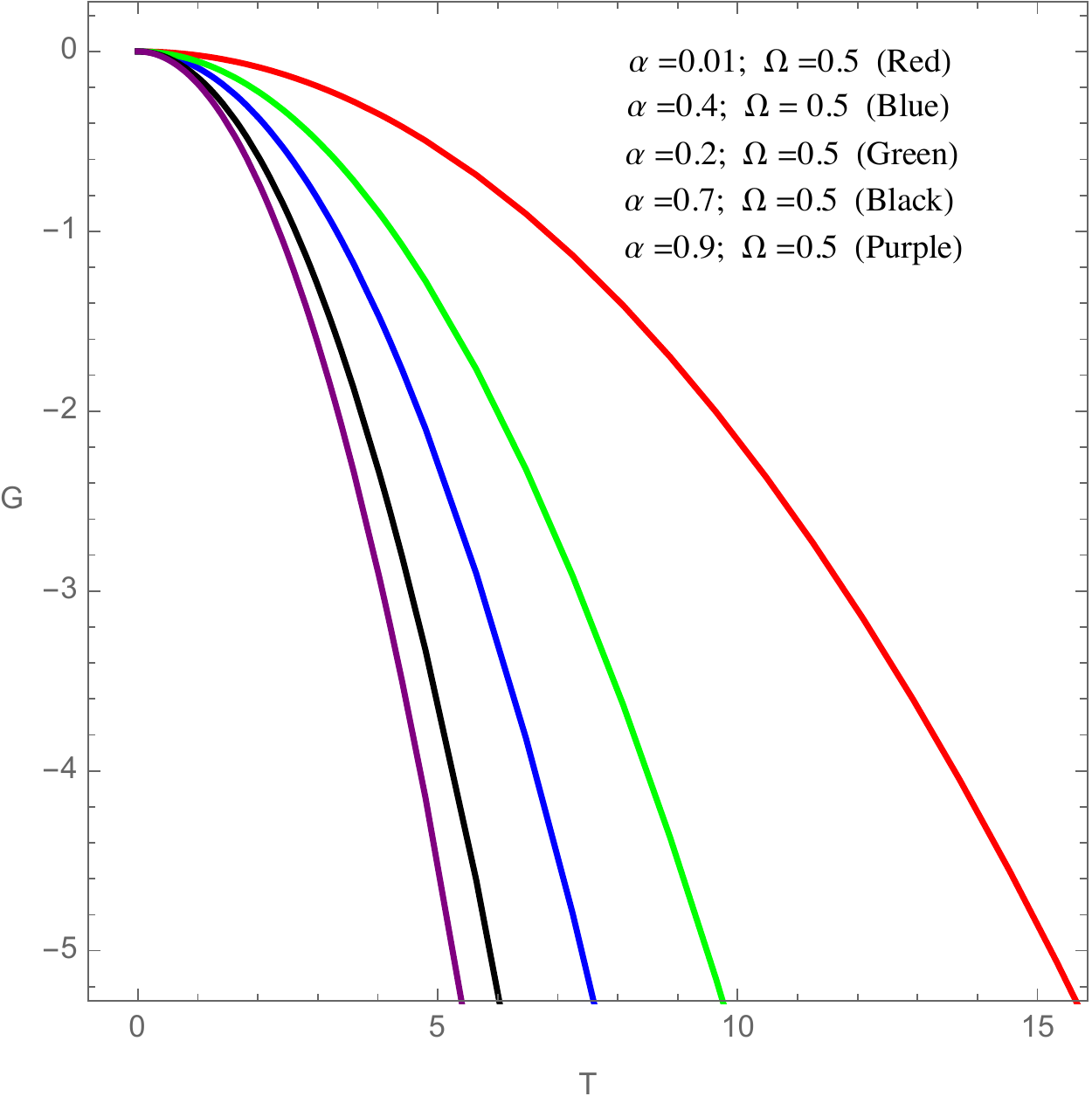}
\caption{    
Gibbs free energy for the \emph{grand-canonical ensemble}:  \emph{Left panel:} Pure exotic (continuous lines) and standard case (dashed lines) for several values of $\Omega$. \emph{Right panel:} different $\alpha$'s and respective $\gamma=1-\alpha$.
}
\label{GibbsOmega}
\end{center}
\end{figure}

In the previous sections, we studied the canonical ensemble with fixed angular momentum but free horizon angular velocity.  Now, we consider the grand-canonical ensemble with the fixed angular velocity of the horizon and free angular momentum.
\begin{equation}
    G_{\Omega}= M - TS - \Omega J = G_{\Omega}(T,P,\Omega)
    \label{eq:GibbsOmega}
\end{equation}
Using Eqs \eqref{TOdef}, we can write $r_{-} = \ell r_{+} \Omega$ and substitute it in the temperature that reads
\begin{equation}
    T=\frac{r_{+} (1 - \ell^2 \Omega^2)}{8 \ell^2 G_{N}}
\end{equation}
and therefore, the temperature is positive if $0< \Omega < 1/\ell$ while $\Omega=1/\ell$ gives the extremal case $T=0$ (the same is true if we express the temperature as function of $r_{-}$). The Gibbs free energy \eqref{eq:GibbsOmega} as function of $T$ and $P$ is
\begin{equation}
    G_{\Omega} = -\frac{\pi  T^2 \left(\sqrt{2 \pi } \gamma  \Omega +4 \pi  \alpha  \sqrt{G_N P}\right)}{8 G_{N}^{3/2} \sqrt{P} \left(8 \pi  G_{N} P-\Omega^2\right)}
\end{equation}
and it is positive iff $\Omega>1/\ell$, that, as we said before, it is not allowed. Consequently, $G_{\Omega}$ will always be negative for $r_{+}>0$.
See the right panel in Figure~\ref{GibbsOmega} for different values of $\alpha$ and $\Omega$. 
\\
In particular, in the exotic limit, the free energy reads
\begin{equation}
    G_{\Omega} (\alpha=0, \gamma=1) = -\frac{\pi ^{3/2} T^2 \Omega}{4 \sqrt{2} G_{N}^{3/2} \sqrt{P} \left(8 \pi  G_{N} P-\Omega^2\right)}
\end{equation}
and in the standard case
\begin{equation}
    G_{\Omega}(\alpha = 1, \gamma=0) = -\frac{\pi ^2 T^2 \sqrt{G_{N} P}}{2 G_{N}^{3/2} \sqrt{P} \left(8 \pi  G_{N} P-\Omega^2\right)}.
\end{equation}
The left panel in Figure~\ref{GibbsOmega} shows the function $G_{\Omega}$ in the exotic and the standard case for different values of $\Omega$.

\section{ Complexity Growth Rate}
Given the interplay between mass and angular momentum that the exotic BTZ black holes show, we wish to check if the bound on complexity growth introduced in \cite{Brown:2015lvg} is valid or violated. 
In the AdS/CFT correspondence, the complexity $\cal{C}$ of a particular state $|\psi \rangle$ can be related to a spacetime region in the bulk. In particular, the complexity=action proposal or the $\cal{C} \cal{A}$-duality states that
\begin{equation} 
   \text{ $\cal{C}$} 
    (|\psi (t_{L}, t_{R}) \rangle)= \frac{\cal{A}}{\pi \hbar}
\end{equation}
where $t_L$ and $t_R$ are, respectively, the time on the left and the right boundary and $\cal{A}$ is the bulk action evaluated on the Wheeler-DeWitt (WdW) patch (see Figure~\ref{fig:WdW_J0}). From this duality, it follows that at late-time the complexity approaches the black hole mass in the following way
\begin{equation}
    \lim_{t_L \rightarrow \infty} \frac{d \cal{C}}{d t_L} = \frac{2M}{\pi \hbar}
\end{equation}
This seems to agree with the Lloyd bound, which limits from above the complexity by the energy
\begin{equation}
    \dot{\cal{C}} \leq \frac{2 E}{\pi \hbar}.
\end{equation}
The connection between the Schwarzschild-AdS WdW patch with the space-time volume and complexity has been pointed out in \cite{Couch:2016exn}. 
It has notably been showed that the thermodynamic volume and the pressure seems to emerge naturally from the late-time rate of growth  (see Appendix \ref{sec:altposs} for more discussions about the connection between thermodynamic volume and complexity).

{In the standard thermodynamic phase-space}, it has been shown \cite{Brown:2015lvg} that  in the case of conserved angular momentum for a BTZ black hole, the complexification bound is 
\begin{equation}
\label{eq:bound}
    \frac{d \cal{C}}{dt} \leq \frac{2}{\pi \hbar} \left[ (M-\Omega J) - (M - \Omega J)_{gs} \right]
\end{equation}
as computed from the Einstein-Hilbert action with Gibbons-Hawking term and corner terms  (see also Appendix \ref{AppendixA} for more details).  The subscript $``gs"$ refers to the ``ground state'', {\it i.e.}  the state of lowest $(M-\Omega J)$ for a given angular momenta. The ground state is there in order to obtain a zero complexification rate for the extremal black hole (that has zero temperature). Therefore the ground state could either be an extremal black hole or empty AdS. 
\\
For the BTZ black hole 
 \begin{equation}\label{eq:MOJ}
    M- \Omega J = \frac{r_+^2-r_-^2}{8 G \ell^2 } = \sqrt{M^2 - \frac{J^2}{\ell^2}}
\end{equation}
which saturates the complexification bound \eqref{eq:bound}  \cite{Brown:2015lvg}. The rate of growth of  the action for the rotating BTZ black hole is
\begin{equation}
    \frac{d \cal{A}}{dt} = \frac{r_{+}^2 - r_{-}^2}{4 G_N \ell^2} = 2 \sqrt{M^2 - \frac{J^2}{\ell^2}}
    \label{eq:ActionBound}
\end{equation}
that shows that the rotating BTZ black hole saturates the bound \eqref{eq:bound}. The ground state for a standard rotating BTZ does not contribute because has $M=0=J$ at fixed $\Omega$. 

In the exotic case the previous bound cannot be  {automatically} valid since for $\alpha<\gamma$ follows $ \ell M < J$.  We shall compute it from the action \eqref{3fE}  and compare it to the corresponding result for the BTZ case, computed using \eqref{3fNI}.
Some of the useful calculations needed to calculate complexity growth for  rotating BTZ black holes can be found in   Appendix \ref{AppendixA}. In particular, using Eq. \eqref{TS} in Eq. \eqref{eq:bound} it is possible to see that, in the standard BTZ case, $d \cal{C}$$/dt\propto TS$  \cite{Couch:2018phr}.  
In the BTZ case one also has 
$d \cal{C}$$/dt\propto P(V_+ - V_-)$ where $V_- = \pi r^2_-$ and $V_+ = \pi r^2_+$ are the  BTZ thermodynamic volumes of the inner horizon and the outer horizon respectively
\cite{Couch:2016exn}. 

%
\subsection{Complexity of the Exotic BTZ Black Hole}
The 
parity-odd  action for $D=3$ Einstein gravity with $\Lambda < 0$   is given by 
\eqref{3fE} \cite{Witten:1988hc}
\begin{equation}
I_{GCS} = \frac{ \ell}{8\pi G}\int_{M^3} \left[  \omega_{a} \wedge \left(  d\omega^{a}+
 {\frac{1}{3}}
\epsilon^{abc}\omega_{b} \wedge \omega_{c}\right)  -\frac{1}{\ell^{2}}
e_{a} \wedge T^{a}\right] 
\label{eq:Spp}
\end{equation}
 Note that contrary to what happens in the case of the EH action, we don't need a boundary term to add to this action in order to have a well defined entropy and variational principle (see, e.g., \cite{Solodukhin:2005ah}); we discuss other possible boundary terms in Appendix \ref{sec:altposs}.
 
The solution to the equations of motion from the action \eqref{eq:Spp} is given by
the space-time triad \cite{Carlip:1994hq}
\begin{eqnarray}
    e^{0} &=& \sqrt{\nu^2 (r) -1} \left( \frac{r_{+}}{\ell} dt - r_{-} d\phi \right) \nonumber\\
    e^{1} &=& \frac{\ell}{\nu(r)} d\left[ \sqrt{\nu^2 (r) -1} \right] = \frac{\ell}{\nu(r)} \left[ \frac{\nu (r) \nu '(r)}{\sqrt{\nu^2 (r)-1}} \right] dr \label{triad}\\
    e^{2} &=& -2 \nu(r) \left( \frac{r_{+}}{\ell^2} dt - \frac{r_{-}}{\ell} d\phi \right) \nonumber
\end{eqnarray}
with $\nu^2 (r):= r^2 - r_{-}^{2} / r_{+}^2 - r_{-}^{2}$ and the prime is the derivative with respect to $r$. The compatible spin connection has the following components:
\begin{eqnarray}
    \omega^{0} &=& -2 \sqrt{\nu^2 (r) -1} \left( \frac{r_{+}}{\ell} d\phi - \frac{r_{-}}{\ell^2} dt \right) \nonumber\\
    \omega^{1} &=& 0 \label{spincon}\\
    \omega^{2} &=& -2 \nu(r) \left( \frac{r_+}{\ell^2} dt - \frac{r_{-}}{\ell} d \phi \right) \nonumber
\end{eqnarray}
and the torsion $T^a = 0$.

It is straightforward to see that the action \eqref{eq:Spp} vanishes for the solution \eqref{triad} (with \eqref{spincon})
since $T^a = 0$ and $\omega^{1} =0$.  The latter implies that $\epsilon^{abc}   \omega_{a} \wedge \omega_{b} \wedge \omega_{c} = 0$. Furthermore, $d\omega^{0} \propto dr\wedge\omega^{0}$ and $d\omega^{2} \propto dr\wedge\omega^{2}$, so the first term in \eqref{eq:Spp} also vanishes.
Therefore  there is no contribution from the action \eqref{eq:Spp} evaluated in the WdW patch $\mathcal{V}$ (see left panel in  Figure~\ref{fig:WdW_J0} and Eq.\eqref{eq:deltaS})   and we are left only with the boundary and the joint contributions from \eqref{genact}; we evaluate these in Appendix \ref{AppendixA}.

Now we consider possible boundary terms for the exotic case. Boundary terms for   Chern-Simons theory have been previously discussed in literature (see for example \cite{Grumiller:2016pqb,Grumiller:2008ie,Arcioni:2002vv, Park:2006gt}. In this particular case of a rotating exotic BTZ black hole, as in the non-exotic case, there is no boundary term in the WDW patch that needs to be evaluated on a space-like surface (see the left panel in Figure \ref{fig:WdW_J0}). Therefore we will be interested here only about the boundary terms that should be evaluated on the null-like surfaces. 

 We know that the variation of the total action ($I_{EH} +  I_{GCS}$) with respect to the metric is well-defined if only the GHY boundary term is included \cite{Solodukhin:2005ah}, up to 
a counterterm ambiguity \cite{Miskovic:2006tm,Miskovic:2009kr,Detournay:2014fva}  that we discuss in Appendix \ref{sec:altposs}. 
    Using this formulation then one can say that there are no other corner terms to  consider, as we know that the corner terms originate from the boundary term in the action \cite{Lehner:2016vdi,Jubb:2016qzt}. Using \eqref{genact}
we   therefore reach the intriguing conclusion that the late-time bound is
\begin{equation}
\label{eq:boundexotic}
    \frac{d \cal{C}}{dt} \leq \frac{\alpha}{\pi \hbar} \frac{r_{+}^2 - r_{-}^2}{4 G_N \ell^2}
\end{equation} 
for the generalized BTZ black hole.  Note that for the pure exotic case $\alpha=0$
the late-time bound vanishes.
 
We have found, therefore, the first example of a $3$-dimensional black hole solution for which the late-time complexity rate growth vanishes. Previous examples have been found only in $2$  and $4$ dimensions \cite{Brown:2018bms,Goto:2018iay}. In those cases,  the inclusion of an additional surface term to the action changes the ensemble and consequently the final result for the complexity growth. In general, the calculation of holographic complexity shows how boundary terms that could be at internal
boundaries become physically meaningful, and the complexity = action conjecture depends on which action one takes into account. In the case of the exotic black hole, the use of complexity=action seems to suggest the need for introducing an additional proper boundary term  (see also the discussion in Appendix \ref{sec:altposs}) that remains to be explored.

\section{Conclusions}
\label{sec:Conclusions}

We have studied the chemistry of generalized exotic BTZ black holes whose thermodynamics are governed by the parameters $\alpha$ and $\gamma$, which serve to swap the roles of mass and angular momentum. The Smarr fomula has been shown to be upheld in all cases.   We find that these generalized exotic black holes have
a number of interesting features, despite the fact that (as with their $\gamma=0$ BTZ counterparts) they exhibit no interesting phase
transition behaviour.

For values of $\alpha > \gamma$, which we term ``majority standard'' black holes, we show there is no particularly interesting behaviour in the thermodynamic quantities, nor the PV diagrams.  However  ``mostly exotic'' black holes, {\it i.e} those for which $\gamma > \alpha$, a variety of interesting behaviour is uncovered.   In contrast to the mostly regular BTZ black holes, we have found that there exists a maximum volume for sufficiently large values of $J \gg T$.   
This phenomenon has not been previously seen, and it means that for given values of $(\alpha,\gamma,\Lambda)$ there is a black hole of maximal size.  Note that this is not necessarily the same as maximal entropy, since the entropy of these black holes is not proportional to their circumference. We also found that the Gibbs free energy is always greater for exotic black holes than for their BTZ counterparts.   

Likewise, we find that reverse isoperimetric inequality  \cite{Cvetic:2010jb} is satisfied for any majority
standard black hole.  However all sufficiently small majority exotic black holes violate this inequality.
 The more exotic and smaller the hole, the greater the violation.  This suggests a connection
between this inequality and the parity properties of the gravitational action, whose  meaning  remains to be understood.  An interesting avenue to explore along these lines is the
recently proposed conjecture \cite{Johnson:2019mdp} between the reverse isoperimetric inequality  and the negativity of the specific heat at constant volume.  Exotic black holes, whose specific heat at constant pressure
is always negative (as shown in figure~\ref{Crp1}) and that always violate the  reverse isoperimetric inequality, will provide an interesting test of this conjecture.

Finally a computation of complexity growth for  exotic BTZ black holes indicates that in general it is smaller by a factor of $\alpha$ as compared to the standard BTZ black hole, as indicated by
\eqref{eq:boundexotic}.  This parameter encodes   the effect of having the gravitational Chern-Simons term in the action.
 Most curiously, the bound on the growth rate vanishes for the purely exotic case
$\gamma = 1$ and $\alpha=0$.   This is because no additional boundary
terms are required if $\gamma \neq 0$, and so the computation is the same as
the BTZ case.  Alternate possibilities are discussed in Appendix  \ref{sec:altposs}.

Our considerations suggest that a theory of quantum gravity including exotic black holes should exhibit
novel properties compared to the usual approaches  \cite{Carlip:1994hq} to the subject.  Any such theory will have to account for their distinct thermodynamic behaviour.  

\section*{Acknowledgements}
 We wish to thank  Daniel Grumiller, Rodrigo Olea, Olivera Miskovic, Jakob Salzer,  Roberto Emparan, 
 Robie Hennigar, Sergey Solodukhin, Hugo Marrochio and Robert C. Myers for interesting discussions and correspondence.
This work was supported in part by the Natural Sciences and Engineering Research Council of Canada.
The work of AMF was supported by a Swiss Government
Excellence Scholarship and is currently supported from ERC Advanced Grant GravBHs692951 and MEC grant FPA2016-76005-C2-2-P. 

\appendix
\section{BTZ black hole complexity growth \label{AppendixA}}
\subsection{General definition of the WdW patch}

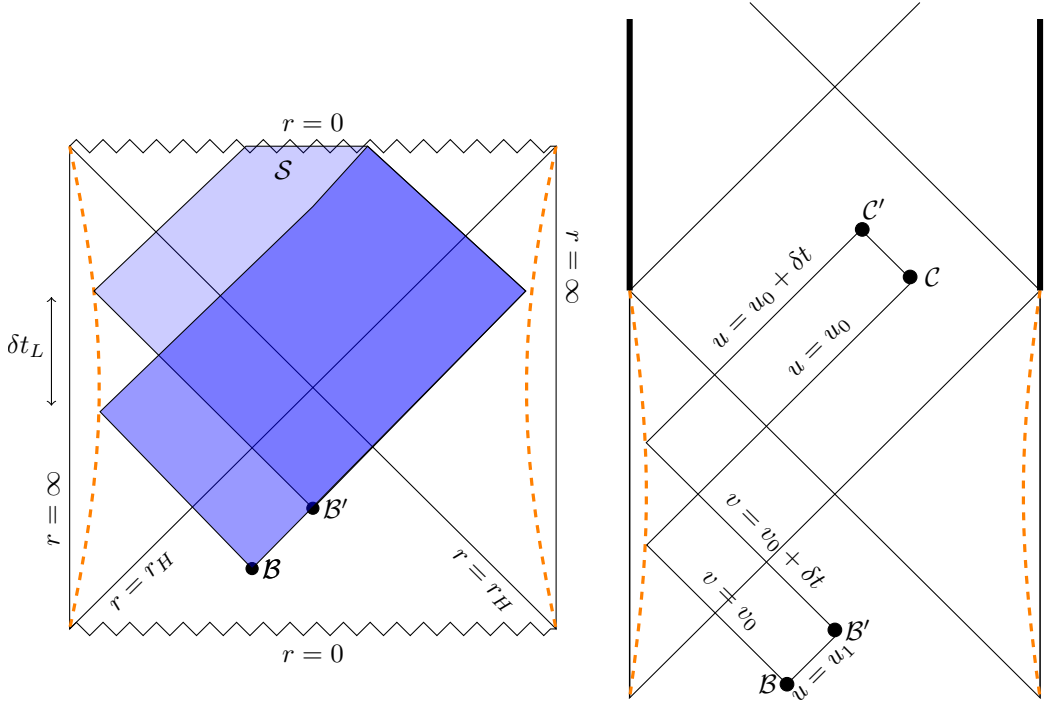
\begin{figure}[h]
%
\begin{center}
%
\begin{tikzpicture}[scale=0.8,font=\small]
\node (I)    at ( 4,0)   {};
\node (II)   at (-4,0)   {};
\node (III)  at (0, 2.5) {}; 
\node (IV)   at (0,-2.5) {}; 
%
\clip (-5,-5.5) rectangle (5,5.5);
\path  
  (II) +(90:4)  coordinate[](IItop)
       +(-90:4) coordinate[](IIbot) 
       +(0:4)   coordinate                  (IIright)
       +(180:4) coordinate 
       (IIleft)
       ;
\draw 
      (IItop) --
      (IIright) -- 
          node[near end, below, sloped] {$r=r_{H}$}
      (IIbot) --
          node[near start, above, sloped] {$r=\infty$}
      (IItop) -- cycle;
\draw[orange,bend left=12,dashed, very thick] (IItop) to (IIbot);
\path 
   (I) +(90:4)  coordinate (Itop)
       +(-90:4) coordinate (Ibot)
       +(180:4) coordinate (Ileft)
       +(0:4)   coordinate (Iright)
       ;
%
\draw  (Ileft) -- (Itop) -- 
    node[near start, above, sloped] {$r=\infty$}
    (Ibot) -- 
    node[near start, below, sloped] {$r=r_{H}$}
    (Ileft) -- cycle;    
\draw[orange,bend right=12,dashed, very thick] (Itop) to (Ibot);    
%
\draw[decorate,decoration=zigzag] (IItop) -- (Itop)
      node[midway, above, inner sep=2mm] {$r=0$};
\draw[decorate,decoration=zigzag] (IIbot) -- (Ibot)
      node[midway, below, inner sep=2mm] {$r=0$};
%
\coordinate[label=right: {$\mathcal{B}$}] (im) at (-1,-3);
\coordinate[draw] (f1) at ($(im)+(-2.5,3) $);
\coordinate[draw] (f2) at ($(f1)+(4,4) $) ;
\coordinate[draw] (f3) at ($(f2)+(3,-3) $) ;
%
%
%
\coordinate[label=right: {$\mathcal{B}^{\prime}$}] (im1) at (-0,-2);
\draw [fill] (-0,-2) circle [radius=0.1];
\coordinate[draw] (f11) at ($(im)+(-2.6,4.6) $);
\coordinate[draw] (f21) at ($(f11)+(2.5,2.4) $) ;
\coordinate[draw] (f31) at ($(f21)+(2,0) $) ;
\coordinate[draw] (f1R) at ($(im)+(4.5,4.6) $);
\draw[fill=blue,fill opacity=0.2] (im1) to (f11) to (f21) to (f31) to (f1R) to (im1);
%
\coordinate[label=right: {$\mathcal{B}$}] (im2) at (-1,-3);
\draw [fill] (-1,-3) circle [radius=0.1];
\coordinate[draw] (f12) at ($(im2)+(-2.5,2.6) $);
\coordinate[draw] (f22) at ($(f12)+(3.5,3.4) $) ;
\coordinate[draw] (f31) at ($(f21)+(2,0) $) ;
\coordinate[draw] (f1R) at ($(im)+(4.5,4.6) $);
\draw[fill=blue,fill opacity=0.4] (im2) to (f12) to (f22) to (f31) to (f1R) to (im2);
%
\coordinate[label=right: {$\delta t_{L}$}] (t) at ($(im1)+(-5.2,2.7) $);
\draw[<->] ($(im1)+(-4.3,3.5) $) -- ($(im1)+(-4.3,1.7) $);
\coordinate[label=right: {$\mathcal{S}$}] (s) at ($(f12)+(2.7,4.0) $) ;
\end{tikzpicture}
%
\begin{tikzpicture}[scale=0.9,font=\small]
%
    \coordinate (s1) at (0,0); 
    \coordinate (s2) at (6,0); 
    \coordinate (PLtop) at (0,4);
    \coordinate (PLbot) at (0,-6);
    \coordinate (PRtop) at (6,4);
    \coordinate (PRbot) at (6,-6);
\draw (s1) -- ++(canvas polar cs:angle=45,radius=6cm); 
\draw (s1) -- ++(canvas polar cs:angle=-45,radius=8.5cm);
\draw (s2) -- ++(canvas polar cs:angle=135,radius=6cm);
\draw (s2) -- ++(canvas polar cs:angle=225,radius=8.5cm);
\draw[line width=0.8mm] (PRtop) -- (s2);
\draw[line width=0.2mm] (s2) -- (PRbot);
\draw[line width=0.8mm] (PLtop) -- (s1);
\draw[line width=0.2mm] (s1) -- (PLbot);
%
\draw[orange,bend right=8,dashed, very thick] (s2) to (PRbot); 
\draw[orange,bend left=8,dashed, very thick] (s1) to (PLbot);
%
\coordinate[label=left: {$\mathcal{B}$}] (B1P) at (2.3,-5.8);
\draw [fill] (B1P) circle [radius=0.1];
\draw (B1P) -- ++(canvas polar cs:angle=45,radius=1.0cm) node[pos=0.5, below, sloped] {$u=u_{1}$};
\draw (B1P) -- ++(canvas polar cs:angle=135,radius=2.9cm) node[pos=0.5, above, sloped] {$v=v_{0}$} -- ++(canvas polar cs:angle=45,radius=5.5cm) node[pos=0.7, above, sloped] {$u=u_{0}$} -- ++(canvas polar cs:angle=135,radius=1.0cm);
\coordinate[label=left: {$\mathcal{C}^{\prime}$}] (C1) at (3.9,1.2);
\draw [fill] (3.4,0.9) circle [radius=0.1];
\coordinate[label=left: {$\mathcal{C}$}] (C1) at (4.7,0.2);
\draw [fill] (4.1,0.2) circle [radius=0.1];
\coordinate[label=right: {$\mathcal{B}^{\prime}$}] (B1) at (3,-5.0);
\draw (B1) -- ++(canvas polar cs:angle=135,radius=3.9cm) node[pos=0.4, above, sloped] {$v=v_{0} + \delta t$} -- ++(canvas polar cs:angle=45,radius=4.5cm) node[pos=0.6, above, sloped] {$u=u_{0} + \delta t$};
\draw [fill] (3,-5.0) circle [radius=0.1];
%
\coordinate (B1PL) at (0.5,-4.0);
\end{tikzpicture}
     \caption{\emph{Left:} WdW patches for a BTZ black hole in the  $J=0$ case.  Note that the time on the right boundary is fixed while the time on the left boundary varies. \emph{Right:} Difference between two WdW patches in the case $J\neq 0$. }
     \label{fig:WdW_J0}
\end{center}
\end{figure}

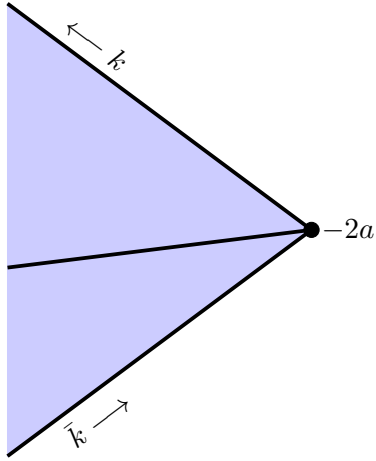
\begin{figure}
    \centering
\begin{tikzpicture}
\coordinate (a) at (0,0);
\coordinate (c) at (4,3);
\coordinate (b) at (0,6);
%
\filldraw[draw=blue!20, fill=blue!20] (a) -- (b) -- (c) -- cycle;
\draw[line width=0.5mm] (c) -- (0,2.5);
\draw[line width=0.5mm] (c) -- (b) node[near end, above, sloped] {$\longleftarrow k$};
\draw[line width=0.5mm] (c) -- (a) node[near end, below, sloped] {$\bar{k} \longrightarrow $};
\draw[black, fill=black]  (c) circle (.1);
\draw (c) node[anchor=west]{$-2a$};
\end{tikzpicture}
    \caption{Composition of null/null joints}
    \label{fig:B-join}
\end{figure}

The WdW patch is a region with boundaries (see for example the shaded area in the left panel in  Figure~\ref{fig:WdW_J0}). 
 In particular, the WdW patch is defined as the spacetime region of a maximally extended black hole enclosed between four light rays originating from the two chosen equal-times $t_{L}$ and $t_{R}$ on each boundary.
Consequently, to evaluate the action in the WdW patch one needs to include both the standard volume integration and the boundary terms
\begin{equation}
    I = \frac{1}{16 \pi G} \int_{M} \sqrt{-g} (R-2 \Lambda) + \frac{1}{8 \pi G} \int_{\partial M} \sqrt{|h|} K.
\end{equation}
where additional corner terms \cite{Booth:2001gx} are essential for
calculating the rate of change of the action when we shift from the WdW patch at coordinate time $t_L$ to the  patch at the time $t_L + \delta t_L$ on the boundary \cite{Lehner:2016vdi}. These ``joint contributions'' are  junction terms where two different boundary surfaces intersect (e.g. see  Figure~\ref{fig:B-join}). 
Including all these latter contributions, the total action to consider in the calculation is:
\begin{equation}
      I =
     \frac{1}{16 \pi G} \left[ \int_{M}  (R-2 \Lambda)\sqrt{-g}\ d^4x +
     2 \int_{\partial M}  K \, \sqrt{|h|} d\Sigma +
     2 \sum_{\mathcal{B}_i}  \text{sign}(\mathcal{B}_i)  \oint a_{\mathcal{B}_i} dS \right]
\end{equation}
where $d\Sigma$ is a volume element in $(D-1)$ and $dS$ is a surface element in $(D-2)$  on the junctions $\mathcal{B}_i$ (see Figure~\ref{fig:WdW_J0}) \cite{Booth:2001gx}.
For example, in the case of the WdW patch for an uncharged, non-rotating black hole, the difference between the actions evaluated in two different WdW patches, $\mathcal{V}_1$ and $\mathcal{V}_2$, is given by  (see left panel in  Figure~\ref{fig:WdW_J0})
\begin{equation} \label{eq:deltaS}
    \delta I_{\mathcal{V}} = I_{\mathcal{V}_1} - I_{\mathcal{V}_2} - \frac{1}{8 \pi G} \int_{\partial M} 
    K d\Sigma + \frac{1}{8 \pi G} \left[ \oint_{\cal{B}^{\prime}} a dS - \oint_{\cal{B}} a dS  \right]
\end{equation}
where the only boundary term is evaluated on the surface $\cal{S}$ close to the singular space-like surface \cite{Lehner:2016vdi}. 
However, in the case of a rotating black hole, there are no space-like boundaries but only null boundaries (see  Figure~\ref{fig:WdW_J0} right panel) \cite{Lehner:2016vdi}. 

In general, the null boundaries can be parametrized in such a way that they do not contribute to the action \cite{Lehner:2016vdi}.

In the case of a rotating BTZ black hole (as for the charged case) there is no contribution from the boundary term on $\partial M$ (see right panel Figure~\ref{fig:WdW_J0})  that is indeed  behind the inner horizon. So we can remove this term from the variation \eqref{eq:deltaS} and we are left only with:
\begin{equation}
    \delta I = I_{\mathcal{V}_1} - I_{\mathcal{V}_{2}} + \delta I_{\mathcal{B},\mathcal{B}^{\prime}} + \delta I_{\mathcal{C},\mathcal{C}^{\prime}} 
\end{equation}
where we used the notation:
\begin{equation}
    \delta I_{\mathcal{B},\mathcal{B}^{\prime}}=\frac{1}{8 \pi G} \left[ \oint_{\cal{B}^{\prime}} a dS - \oint_{\cal{B}} a dS  \right].
\end{equation}
The analysis presented in \cite{Lehner:2016vdi} provides a meaningful comparison of the action between different WdW patches.  An extended discussion about possible divergent contributions to the action is presented in \cite{Carmi:2016wjl}.

\subsection{Rotating BTZ in Einstein-Cartan formalism}

Let us recap the procedure presented for the rotating BTZ black hole. 
The standard (parity-even) Einstein AdS action in $D=3$ is (now we write also the explicit wedge
products among forms)
\begin{equation}
S_E = \frac{1}{8\pi G} \int_{M^3} \left[  e_{a} \wedge R^{a}-\frac{1}{6\ell^{2}}\epsilon^{abc}%
e_{a} \wedge e_{b} \wedge e_{c}\right]  
\label{3fN}
\end{equation}
where instead of working with the metric $g_{\mu \nu}$, we are working with the auxiliary \emph{frame fields}, or vielbein, $e^{a}_{\mu}$ with $a=0,1,2$ and defined by the relation with the metric
\begin{equation}
    g_{\mu \nu} (x) = e^{a}_{\mu}(x) e^{b}_{\nu}(x) \eta_{a b}
\end{equation}
with $\eta_{a b}$ the metric of the flat 3D Minkowski spacetime.
The vielbein define a basis in the space of the differential forms. 
In the ``dual notation'' (valid only in $D=3$) that we used in $S_E$, we have
\begin{equation}
    R_{a} = \frac{1}{2} \epsilon_{abc} R^{bc} \leftrightarrow
    R^{ab}= - \epsilon^{abc} R_c
\end{equation}
therefore, Eq. \eqref{3fN} can be written:
\begin{equation}
S_E = \frac{1}{16\pi G} \int_{M^3} \epsilon_{abc} \left(  e^{a} \wedge 
 R^{bc} 
-\frac{1}{3\ell^{2}}%
e^{a} \wedge e^{b} \wedge e^{c}\right) .
\label{3fN-1}
\end{equation}
Since, by definition, one has $ e^{a} = e^{a}_{\rho} dx^{\rho}$ and
\begin{equation}
    R^{bc} =\frac{1}{2} R^{bc}_{\; \mu \nu}(x) dx^{\mu} \wedge dx^{\nu} = \frac{1}{2} \left(e^{b}_{\beta} e^{c}_{\gamma} R^{\beta \gamma}_{\; \mu \nu} \right) dx^{\mu} \wedge dx^{\nu}
\end{equation}
then, Eq. \eqref{3fN-1} corresponds to 
\begin{equation}
S_E = \frac{1}{16\pi G} \int_{M^3} \epsilon_{abc} \left[ 
\frac{1}{2} \left( e^{a}_{\rho}
 e^{b}_{\beta} e^{c}_{\gamma} R^{\beta \gamma}_{\; \mu \nu} \right) 
-\frac{1}{3\ell^{2}}%
e^{a}_{\rho} e^{b}_{\mu} e^{c}_{\nu}  \right] dx^{\rho} \wedge dx^{\mu} \wedge dx^{\nu}. 
\end{equation}
Using the Levi-Civita in frame components
\begin{equation}
    \epsilon_{\mu \nu \rho } = e^{-1} \epsilon_{a b c} e^{a}_{\mu} e^{b}_{\nu} e^{c}_{\rho}
\end{equation}
with $e=\text{det} e^{a}_{\mu} = \sqrt{-g}$, one gets
\begin{eqnarray}
   S_E &=& \frac{1}{16\pi G} \int_{M^3} e  \left( 
\frac{1}{2} \epsilon_{\rho \beta \gamma} R^{\beta \gamma}_{\; \mu \nu}  
-\frac{1}{3\ell^{2}}%
\epsilon_{\rho \mu \nu}  \right) dx^{\rho} \wedge dx^{\mu} \wedge dx^{\nu}\\
 &=& \frac{1}{16\pi G} \int_{M^3} d^3x \sqrt{-g}  \left( 
\frac{1}{2} \epsilon_{\rho \beta \gamma} \epsilon^{\rho \mu \nu} R^{\beta \gamma}_{\; \mu \nu}  
-\frac{1}{3\ell^{2}}%
\epsilon_{\rho \mu \nu} \epsilon^{\rho \mu \nu}  \right)
\end{eqnarray}
where in the last line we have used $dx^{\rho}\wedge dx^{\mu} \wedge dx^{\nu} = \epsilon^{\rho \mu \nu} d^{3}x$.
Finally, using 
the following equalities
\begin{eqnarray}
   \frac{1}{2} \epsilon_{\rho \beta \gamma} \epsilon^{\rho \mu \nu} R^{\beta \gamma}_{\; \mu \nu} & = &
   \frac{1}{2} \left( \delta_{\beta}^{\mu} \delta_{\gamma}^{\nu} - \delta_{\beta}^{\nu} \delta_{\gamma}^{\nu} \right)R^{\beta \gamma}_{\; \mu \nu} = \frac{1}{2} \left( R^{\mu \nu}_{\; \mu \nu} - R^{\nu \mu}_{\mu \nu} \right) = R\\
   \epsilon_{\rho \mu \nu} \epsilon^{\rho \mu \nu} & = & - 3!
\end{eqnarray}
we get
\begin{equation}
    I_{E} = \frac{1}{16 \pi G} \int_{M^3} d^3x \sqrt{-g} \left( R - 2 \Lambda \right)
\end{equation}
with negative cosmological constant $\Lambda$
\begin{equation}
    \Lambda= - \frac{(D-2)(D-1)}{2 \ell^2} = - 1/\ell^2 
\end{equation}
From this action, we can then proceed in the standard way and get the following contributions to the variation of the action in the WdW patch.
\subsection{Volume contributions to the WdW patch}

In the presence of a negative cosmological constant, the solution to the equations of motion for the action  \eqref{3fNI} indicates that the Ricci scalar is a constant, $R=6\Lambda$. Hence the volume contribution to the evolution of the WdW patch is  given by
\begin{equation}
     \delta I_\mathcal{V}  = I_{\mathcal{V}_1} - I_{\mathcal{V}_2} = - \frac{4}{16 \pi G \ell^2} \int \sqrt{-g}\; d^3 x
\end{equation}
Writing this integral explicitly for the WdW patchs for the rotating black hole in  Figure~\ref{fig:WdW_J0} (right panel) and considering late times, we obtain 
\begin{eqnarray}
    \delta I_{\mathcal{V}} 
    & = & - \frac{1}{4 \pi G \ell^2} \int_{r_{-}}^{r^{+}} r\; dr \int_0^{2\pi} \;d\theta\; \delta t \nonumber\\
    & = & - \left[ \left( m - \Omega_{+} j \right) - \left( m - \Omega_{-} j \right) \right]\; \delta t \nonumber \\
    & = & 
     - \frac{r_{+}^2 - r_{-}^2}{4 G  \ell^2 }\; \delta t
\label{btzvolcontr}     
\end{eqnarray}
with 
\begin{equation}
    m = \frac{r_{+}^2 + r_{-}^2}{8G \ell^2}, \quad j=\frac{r_{+}r_{-}}{4 G \ell},  \quad  \Omega_{+}=\frac{4Gj}{r_{+}^2}, \quad \Omega_{-}= \frac{4Gj}{r_{-}^2}
    \label{eq:jANDm}
\end{equation}
and where we could write the last term as $ \delta I_{\mathcal{V}}  = - TS \delta t$, since
\begin{equation}\label{TS}
    T= \frac{r_{+}^2 - r_{-}^2}{2 \pi \ell^2 r_{+}}, \quad S= \frac{\pi r_{+}}{2G}
\end{equation}
While the right-hand side of \eqref{btzvolcontr}  is proportional to  the result \eqref{eq:ActionBound},  we still need to add the joint (or corner) terms to get the final result.  We also note that the rotating BTZ black hole is a special case for which $(m - \Omega_{+} j) = -  (m - \Omega_{-} j)$ \cite{Cai:2016xho,Brown:2015lvg}.  

\subsection{Joint contributions}

Now we study the joint terms. It is useful to write 
\begin{eqnarray}
g_{\mu\nu}=\left(\begin{array}{ccc}
r^{2}(N^{\phi})^2-N^{2} & 0 & r^{2}N^{\phi}\\
0 & N^{-2} & 0\\
r^{2}N^{\phi} & 0 & r^{2} 
\end{array}\right) \qquad 
 g^{\mu\nu}=\left(\begin{array}{ccc}
-N^{-2} & 0 & N^{\phi}/N^{2}\\
0 & N^{2} & 0\\
N^{\phi}/N^{2} & 0 & \frac{1}{r^{2}}-\frac{(N^{\phi})^2}{N^{2}}
\end{array}\right)
\label{downupmetrics}
\end{eqnarray}
which follow from the rotating BTZ black hole metric  \eqref{BTZ1}. 
 We want to evaluate the joint terms
$
\pm2\ointop_{\mathcal{B}}a dS
$
at the intersection between two null-segments, $\mathcal{B}$ and
$\mathcal{B}^{\prime}$ that are in this case one dimensional surfaces.
The area element in $\mathcal{B}$ and $\mathcal{B}^{\prime}$ is
$dS=rd\Omega^{1}=r\ d\phi$ and 
\begin{equation}
a=\ln\left(-\frac{1}{2}k\cdot\bar{k}\right)
\label{eq:adS}
\end{equation}
where 
$k_{a}$ and $\bar{k}_{a}$ are the normals to the null surfaces respectively
to $v=v_{0}$ and $u=u_{1}$ that meet in $\mathcal{B}$ (or to $v=v_{0}+\delta t$
in $\mathcal{B}^{\prime}$, see  Figure~\ref{fig:WdW_J0}). 
We can choose the two normals as the
affinely parametrized vectors defined by the gradient of the implicit equations of the
surfaces \cite{Lehner:2016vdi}:
\begin{eqnarray}
\label{ka-kabar}
k_{a} &=& -c \,\partial_{\alpha}\left(v-v_{0}\right)=-c \,\partial_{\alpha}\left(t-r^{*}\right)\\
\overline{k}_{a} &=& \overline{c} \,\partial_{\alpha}\left(u-u_{1}\right)=\overline{c} \,\partial_{\alpha}\left(t+r^{*}\right)
\end{eqnarray}
where $c$ and $\overline{c}$ are normalization constants 
 that define the asymptotic normalization $k \cdot \hat{t}_{L} = - c$
and $\bar{k} \cdot \hat{t}_{R} = - \bar{c}$, where $\hat{t}_{L}=\partial_{\hat{t}_{L}}$ is the asymptotic Killing vector describing the time flow in the left boundary theory and $\hat{t}_R=\partial_{t_{R}}$ is the corresponding vector for the right boundary theory,  $v_{0}$ and $u_{1}$ are constants and $v:=t-r^{*}$,
$u:=t+r^{*}$. 
Recalling that $r^{*}=\int g_{11}dr=\int\frac{1}{N^{2}}dr$
we can write the non zero components of the normals:
\begin{eqnarray*}
k_{0} &=& -c, \quad k_{1}=cN^{-2} \qquad
\overline{k}_{0} = \overline{c}, \qquad \overline{k}_{1}=\overline{c}N^{-2}
\end{eqnarray*}
so, using this result and the inverse metric, the scalar product in Eq. \eqref{eq:adS} is
\begin{eqnarray*}
k\cdot\overline{k} & = & g^{00}k_{0}\overline{k}_{0}+g^{11}k_{1}\overline{k}_{1} = \frac{2}{N^{2}}\left(c\overline{c}\right)
\end{eqnarray*}
so $a=\ln\left(-c\overline{c}/N^{2}\right)$. We can now evaluate
the contribution of the joint terms
\begin{equation}
   (16\pi G) \;   \delta I_{\mathcal{B},\mathcal{B}^{\prime}}
   = 2 \left[ \oint_{\cal{B}^{\prime}} a \;dS - \oint_{\cal{B}} a\; dS  \right] = 
2 \int_{0}^{2\pi}d\phi \left.\left[\ln\left(-\frac{c\overline{c}}{N^{2}}\right)r\right]\right|^{r=r_{\mathcal{B}}}_{r=r_{\mathcal{B}^\prime}   }
 \label{eq:hdrB}
\end{equation}

Since $t=\frac{1}{2}\left(u+v\right)$ and $r^{*}=\frac{1}{2}\left(u-v\right)$, 
for $\mathcal{B}$ and $\mathcal{B}^{\prime}$ we have
\begin{eqnarray}
\begin{array}{c}
\!\!\!\!\!\!\!\!\!\!\!\!\!\mathcal{B}\equiv\left(v=v_{0},u=u_{1}\right)\\
\mathcal{B^{\prime}}\equiv\left(v=v_{0}+\delta t,u=u_{1}\right)
\end{array}\Rightarrow\begin{array}{c}
r_{\mathcal{B}}^{*}=\frac{1}{2}\left(u_{1}-v_{0}\right)\\
\qquad r_{\mathcal{B^{\prime}}}^{*}=\frac{1}{2}\left(u_{1}-v_{0}-\delta t\right)
\end{array}
\end{eqnarray}
and so for small $\delta t$ 
\begin{eqnarray}
r_{\mathcal{B^{\prime}}}=r\left(r_{\mathcal{B^{\prime}}}^{*}\right)=r\left(r_{\mathcal{B}}^{*}-\frac{1}{2}\delta t\right)\cong r\left(r_{\mathcal{B}}^{*}\right)-\frac{1}{2}\left.\frac{dr}{dr^{*}}\right|_{r^{*}=r_{\mathcal{B}}^{*}}\!\!\!\!\!\!\!\!\!\!\!\! \delta t
\quad = r\left(r_{\mathcal{B}}^{*}\right)-\frac{1}{2}\left. N^{2}\right|_{r^{*}=r_{\mathcal{B}}^{*}}  \delta t
\end{eqnarray}
Defining $h(r):= r\ln\left(-c\overline{c}N^{-2}\right)$  we have \cite{Lehner:2016vdi}
\begin{equation}
4\pi\left[h\left(r_{\mathcal{B}^{\prime}}\right)-h\left(r_{\mathcal{B}}\right)\right]
  = 4\pi\frac{1}{2}\left[r\ \frac{dN^{2}}{dr}+N^{2}\ln\left(-\frac{c\overline{c}}{N^{2}}\right)\right]\delta t
\end{equation}
where for late times, $r_{\mathcal{B}}\rightarrow r_{+}$ and  so $N\rightarrow0$. 
In this limit, the logarithmic 
contribution vanishes and the final expression is 
\begin{eqnarray}
    \delta I_{\mathcal{B},\mathcal{B}^{\prime}} &=& \frac{2 \pi}{16 \pi G} \left.\left[ r \left( \frac{2 r}{\ell^2} - \frac{32 G^2 j^2}{ r^{3}} \right) \right] \right|_{r_{+}} 
    \delta t = \frac{1}{8 G} \left( \frac{2 r_{+}^2}{\ell^2} - \frac{2 r_{-}^2}{\ell^2} \right) \delta t
\end{eqnarray}
for a rotating BTZ black hole.  The same calculation can be done for the junctions $\mathcal{C}$ and $\mathcal{C}^{\prime}$ (see  Figure~\ref{fig:WdW_J0}), yielding
\begin{eqnarray}
    \delta I_{\mathcal{C},\mathcal{C}^{\prime}} &=&
    - \frac{2 \pi}{16 \pi G} \left.\left[ r \left( \frac{2 r}{\ell^2} - \frac{32 G^2 j^2}{ r^{3}} \right) \right] \right|_{r_{-}} \delta t = 
    - \frac{1}{8 G} \left( \frac{2 r_{-}^2}{\ell^2} - \frac{2 r_{+}^2}{\ell^2} \right) \delta t
\end{eqnarray}
where we used \eqref{eq:jANDm}.  Hence
\begin{equation}
    \delta I_{\mathcal{B},\mathcal{B}^{\prime}} + \delta I_{\mathcal{C},\mathcal{C}^{\prime}} = \frac{r_{+}^2 - r_{-}^2}{2 G \ell^2} \delta t
    \label{eq:join-sum}
\end{equation}

All together, we find 
\begin{eqnarray}
     \delta I_\mathcal{V} +   \delta I_{\mathcal{B},\mathcal{B}^{\prime}} + \delta I_{\mathcal{C},\mathcal{C}^{\prime}}
    & = &  \frac{ \left( r_{+}^2 - r_{-}^2 \right) }{4G\ell^2}  \delta t 
    \label{btzcomp}    
\end{eqnarray}
which can also be written as
\begin{eqnarray}
     \delta I_\mathcal{V} +   \delta I_{\mathcal{B},\mathcal{B}^{\prime}} + \delta I_{\mathcal{C},\mathcal{C}^{\prime}}
    & = & 2 \left[ \left( m - \Omega_{+} j \right) - \left( m - \Omega_{-} j \right) \right] \delta t \nonumber\\
    &=&2 \left[ \left( M - \Omega_{+} J \right) - \left( M - \Omega_{-} J \right) \right] \delta t
\label{btzcomp2}    
\end{eqnarray}
using \eqref{SystM} and \eqref{SystJ} with $\gamma=0$.  This is commensurate
with \eqref{eq:ActionBound}, previously obtained \cite{Cai:2016xho,Brown:2015lvg} via a different method.
There is likewise a similar result for a non-rotating BTZ black hole and different regularization procedure \cite{Carmi:2017jqz}, and a generalization of this to topological geons has likewise been recently obtained 
\cite{Sinamuli:2018jhm}.

\subsection{Boundary counterterm for rotating BTZ}

Here we calculate the boundary counterterms that are necessary to render the action reparametrization invariant for the null boundaries of the WDW patch. These boundary counterterms play a relevant role in the calculation of complexity growth \cite{Agon:2018zso, Caceres:2018luq}. For  a null hypersurface 
they in $(2+1)$ dimensions the counterterm contribution to the action is
  \cite{Lehner:2016vdi, Reynolds:2016rvl}  
  \begin{eqnarray}
  I_{\text{boundary}} = \frac{1}{8 \pi G} \int_{\Sigma}  d\lambda dr \; \sqrt{\gamma}\; \Theta \log \left( \Tilde{L} \Theta \right)
\end{eqnarray}
where $\gamma$ is the induced metric on the null sheet of the WDW patch, $\Theta$ is its expansion given by $\Theta := \partial_{\lambda} \log \sqrt{\gamma}$ and $\tilde{L}$ is an arbitrary constant.

Now we calculate the metric $\gamma$ for the rotating BTZ black hole using the procedure illustrated in \cite{Balushi:2019pvr}. Using the  rotating-BTZ metric \eqref{BTZ1} we 
obtain the tortoise coordinate $r_{*}$ that satisfies
\begin{equation}
    g^{\alpha \beta} \partial_{\alpha} v \partial_{\beta} v = g^{tt} + g^{rr} (d_r r_{\star}(r))^2  = 0
\end{equation}
that implies 
\begin{equation}
    \left( \frac{dr_{*}}{dr} \right)^2 = - \frac{g^{tt}}{g^{rr}} = \frac{1}{N^{4}} \rightarrow dr_{*} = \frac{dr}{N^2}
 \end{equation}
transforming the metric \eqref{BTZ1} to
 \begin{equation}
 \label{eq:metr-BTZ-J-2}
     ds^2 = N^{2} (-dt^2 + dr_{*}^2) + r^2 (d\phi - N^{\phi} dt)^2.
 \end{equation}
where  $N$ and  $N^{\phi} $ are given in \eqref{lpseshft}.

Further employing the transformations 
\be 
    d(t-r_{*}) = du  =  -\frac{dU}{\kappa U} \qquad 
    d(t+ r_{*}) = dv  =  \frac{dV}{ \kappa V}.
\ee
yields 
\begin{equation}
\label{eq:metr-BTZ-J-3}
    ds^2 = N^2 \left( \frac{dU}{\kappa U} \frac{dV}{\kappa V} \right) + r^2 \left[d\phi -\frac{N^{\phi}}{2 \kappa} \left( -\frac{dU}{U} + \frac{dV}{V} \right) \right]^2
\end{equation}
from \eqref{eq:metr-BTZ-J-2}.

Although the first term in \eqref{eq:metr-BTZ-J-3} is regular at the horizon, the 2nd term  is not regular at either of $U=0$ or $V=0$.  This problem can be dealt with by 
defining \cite{Balushi:2019pvr, Chan:1994rs}
\begin{equation}
    \varphi_{\pm} = \phi \pm \int \frac{N^{\phi}}{N^2} \; dr,  
  \end{equation}
yielding  for  \eqref{eq:metr-BTZ-J-3} 
\begin{equation}
    ds^2 = N^2 \left( \frac{dU}{\kappa U} \frac{dV}{\kappa V} \right) + r^2 \left[ d\varphi_{+} - N^{\phi} dv \right]^2
\end{equation}
valid at $U=0$ and 
\begin{equation}
    ds^2 = N^2 \left( \frac{dU}{\kappa U} \frac{dV}{\kappa V} \right) + r^2 \left[ d \varphi_{-} - N^{\phi} du \right]^2.
\end{equation}
valid at $V=0$.  The induced metric is given
\begin{eqnarray}
    dx^{A}dx^{B} \gamma_{AB} &=& ds^2(U=\text{const}) = r^2 \left[ d\varphi_{+} - N^{\phi} dv \right]^2\\
    dx^{A}dx^{B} \gamma_{AB} &=& ds^2(V=\text{const}) = r^2 \left[ d\varphi_{-} - N^{\phi} du \right]^2
\end{eqnarray}
and so the determinant is $\gamma = r^2$.   

Using  the affine parameter $\lambda = r/\alpha$,  the expansion parameter $\Theta$ is  
\begin{equation}
    \Theta = \partial_{\lambda} \log (\sqrt{\gamma})  = \frac{\alpha}{r}
\end{equation}
which is the same counterterm as that for the non-rotating-BTZ black hole.  Hence
\begin{eqnarray}
  \Delta I_{\Sigma}^{BTZ} &=& \frac{\Omega_{k,1}}{4 \pi G} 
  \int^{r_{\text{max}}}_{0} dr\; \log \left( \frac{\alpha \Tilde{L}}{r}  \right) + \frac{\Omega_{k,1}}{4 \pi G}
  \int^{r_{\text{max}}}_{r_{\text{m}}} dr\; \log \left( \frac{\alpha \Tilde{L}}{r} \right) \nonumber\\
  &=& \frac{1}{G} r_{\text{max}} \left( \log \frac{r_{\text{max}}}{\alpha \Tilde{L}} - 1 \right) + \frac{1}{2 G} r_{\text{m}} \left( \log \frac{r_{\text{m}}}{\alpha \Tilde{L}} -1 \right)
\label{ctcontrib}  
\end{eqnarray}
considering that there are two past null boundaries and two future null boundaries in total.

The quantity  $r_{\text{max}}$ is the
position of the UV regulator surface  {(at the boundary)}, and  $r_{m}$ is the location of the
intersection of the two past null boundaries; we shall set $r_{m}= 0$ when these boundaries end on the past singularity.

However the causal structure of the rotating BTZ black hole is similar to the charged AdS black hole: there are inner and outer horizons, and the WDW patch does not end on the singularity.  Specifically  $r_\text{m}$ never ends on the singularity (see right panel in Fig. \ref{fig:WdW_J0}); rather the intersection points are at  $r(\mathcal{B}), r(\mathcal{B^{\prime}}),r(\mathcal{C}),r(\mathcal{C^{\prime}})$.

Since the counterterm for the $AdS_{3}$ is  
\begin{equation}
    \Delta I_{\Sigma}^{AdS} = - \frac{1}{2 G} r_{\text{max}} \left( \log\frac{r_{\text{max}}}{\alpha \Tilde{L}} -1 \right)
    \end{equation}
we obtain from   \eqref{ctcontrib}  the total regularized counterterm
\begin{equation}
    \Delta I^{reg} = \Delta I_{\Sigma}^{BTZ} - 2  \Delta I_{\Sigma}^{AdS} =  \frac{1}{2 G} r_{\text{m}} \left( \log \frac{r_{\text{m}}}{\alpha \Tilde{L}} -1 \right).
\end{equation}
We can show that adding this term to the total action cancels the dependence on $\alpha$ (i.e. is independent of the normalization constant $\alpha$ appearing in the null normals).
If we adopt for this subsection the normalization of the null vectors in \eqref{ka-kabar} to be the same:
\begin{equation}
    k \cdot \hat{t}_{L} = \bar{k} \cdot \hat{t}_{R} = \pm \alpha
\end{equation}
where the sign $+$ is for the future null surfaces and the sign $-$ is for the past null surfaces,
the join contribution \eqref{eq:hdrB} is 
\begin{equation}
    I_{jnt} = \frac{1}{8 \pi G} \int_{0}^{2 \pi} d\phi\, r_{\text{m}} \log  \left( \frac{\alpha^2}{N^2 (r_{m})} \right) = \frac{1}{4 G} r_{\text{m}} \log  \left( \frac{\alpha^2}{N^2 (r_{m})} \right)
 \end{equation}
 therefore, if we sum 
 \begin{eqnarray}
     \Delta I^{reg} + I_{jnt} &=& \frac{r_{m}}{2G} \left[ \log\frac{r_{\text{m}}}{\alpha \Tilde{L}} - 1 + \frac{1}{2} \log\frac{\alpha^2}{N^2 (r_{\text{m}})}  \right]\\
     &=& \frac{r_{\text{m}}}{2G} \left[ \log \frac{r_{\text{m}}}{\alpha \Tilde{L}} + \log\frac{\alpha}{N(r_{\text{m}})} -1 \right]\\
     &=& \frac{r_{\text{m}}}{2 G} \left[\log\frac{r_{\text{m}}}{\Tilde{L} N(r_{\text{m}})} -1  \right]
 \end{eqnarray}
 we find that the result is independent of the normalization
constant present in the null normal normalization.

\section{Alternate Possibilities}
\label{sec:altposs} 
 
The preceding calculation {assumed that the boundary term in the action \eqref{genact} was a functional of the boundary metric and unit normal} only.  Using
 the tetrad formalism, it is in fact possible to see that the variation of the various terms in the 
action \eqref{genact} with respect to $e_{a}$ and $\omega_{a}$ give the following boundary terms:
\begin{eqnarray}
    \delta_{\omega} I_{EH} &=& \int d \left( e_{a} \wedge \delta \omega^{a} \right) \label{eq:BT-1}
    \label{eq:b_oEH}\\
    \delta_{e} I_{GCS} &=& \int d \left( e_{a} \wedge \delta e^{a} \right) 
    \label{eq:b_eGCS}\\
    \delta_{\omega} I_{GCS} &=& \int d \left( \omega_{a} \wedge \delta \omega^{a} \right) 
    \label{eq:b_oGCS}
\end{eqnarray}
and while Eq. \eqref{eq:BT-1} gives the well-known GHY term and the second  \eqref{eq:b_eGCS} can be canceled using the boundary condition for $e^{a}$,  a boundary term that could cancel
\eqref{eq:b_oGCS}  could play a role in the calculation of the complexity growth.  However no such boundary term exists.  Furthermore, a Fefferman-Graham coordinate expansion indicates that it depends
only on variations of the leading part of the asymptotic metric \cite{Kraus:2005zm}, and so vanishes in spacetimes having asymptotically AdS boundary conditions.
Note also that since the quantity $\sqrt{|h|}K -2\sqrt{|h|}/\ell$ is a topological invariant 
of the background metric in $(2+1)$ dimensions
\cite{Miskovic:2006tm,Miskovic:2009kr,Detournay:2014fva},  there is some ambiguity
in choosing the boundary term in \eqref{genact}; this does not affect the computations in
the previous appendix.

Another hint that something interesting can happen in this case is if one wants to analyze the connection between the thermodynamic volume and the complexity growth given in \cite{Couch:2016exn}. It has been proposed indeed that the complexification rate could be proportional to the product $PV$ (this is called ``Complexity=Volume" 2.0). This is because, the complexity growth should be be proportional to the number of degrees of freedom that are in the boundary CFT given by the central charge multiplied for the number of lattice sites $n$ (roughly the volume divided by the lattice spacing) multiplied by the number of circuit steps $n_c$.  

 In $D$ spacetime dimensions (setting $G_N=1$), the central charge scales as $c  \sim \ell^{D-2} \sim P^{-(D-2)/2}$,
and the number of sites $n \sim  V/\epsilon^{D-1}$, where the lattice spacing $\epsilon \sim  \ell\sim  \sqrt{\frac{1}{P}}$.  The number of circuit steps $n_c \sim \Delta t /\epsilon$, so 
\be
 \mathcal{C} \propto  \frac{1}{P^{(D-2)/2}} \times {V}{P^{(D-1)/2}}\times  \Delta t \sqrt{P} =  PV  \Delta t 
\ee
up to some normalization\footnote{We are grateful to Robie Hennigar for clarification of this argument.}.
In the standard bulk BTZ case  the Brown-Henneaux formula  
\begin{equation}\label{c-charge}
    c= \frac{3 \ell}{2 G_{N}} \propto \sqrt{\frac{1}{P}}
\end{equation}
suggests that one can write \cite{Couch:2016exn}
\begin{eqnarray}\label{C-PV}
 \mathcal{C}  =  P V (t_{L} + t_{R})
\end{eqnarray}
for $D=3$.   
However, if we add the GCS term to the action, it has been shown that the boundary CFT has different values of the central charge \cite{Solodukhin:2005ah, Townsend:2013ela}
\be
    c_{L}= \frac{3}{2} \frac{(\alpha - \gamma )\ell}{G_{N}}, \qquad  c_{R}= \frac{3}{2} \frac{(\alpha  + \gamma )\ell}{G_{N}}.
\ee
Now, in order to obtain the well known result for the BTZ black hole, we need to define the quantity $c_{+} = (c_{L} + c_{R})/2$. This quantity has been derived in  \cite{Solodukhin:2005ah} using the conformal anomaly to determine the covariantly conserved stress-energy tensor in the presence of the gravitational Chern-Simons term in the action. Using $c_{+}$, we obtain the correct result both in the standard case ($\gamma=0$), when $c_{L} = c_{R}$ and the quantity $c_{+}$ is equal to~\eqref{c-charge} giving the result~\eqref{C-PV}, and for the more general exotic case, our final result Eq.~\eqref{eq:boundexotic}. 

\bibliographystyle{JHEP}
\bibliography{LBIB0_utf1}

\end{document}